\documentclass[conference]{IEEEtran}
\IEEEoverridecommandlockouts
\usepackage{cite}
\usepackage{amsmath,amssymb,amsfonts}
\usepackage{algorithmic}
\usepackage{graphicx}
\usepackage{textcomp}
\usepackage{xcolor}

\usepackage{subfigure}
\usepackage{times}
\usepackage{helvet}
\usepackage{courier}
\usepackage{todonotes}
\usepackage{booktabs}
\usepackage{graphicx}

\usepackage{tabularx}
\usepackage{multirow}
\usepackage{caption}
\usepackage{diagbox }
    \usepackage{array}
\newcolumntype{J}{>{\centering\arraybackslash}m{12cm}}

    \usepackage{array}
\newcolumntype{L}{>{\centering\arraybackslash}m{7.5cm}}

\usepackage{array}
\newcolumntype{j}{>{\centering\arraybackslash}m{4cm}}

\makeatletter
\newcommand{\linebreakand}{%
  \end{@IEEEauthorhalign}
  \hfill\mbox{}\par
  \mbox{}\hfill\begin{@IEEEauthorhalign}
}
\makeatother

\makeatletter
\def\thickhline{%
  \noalign{\ifnum0=`}\fi\hrule \@height \thickarrayrulewidth \futurelet
   \reserved@a\@xthickhline}
\def\@xthickhline{\ifx\reserved@a\thickhline
               \vskip\doublerulesep
               \vskip-\thickarrayrulewidth
             \fi
      \ifnum0=`{\fi}}
\makeatother

\newlength{\thickarrayrulewidth}
\newcommand{\data}{{\sf {MM-COVID}}}
\begin{document}
 \title{{\data}: A Multilingual and Multimodal Data Repository for Combating COVID-19 Disinformation
}

\author{\IEEEauthorblockN{Yichuan Li }
\IEEEauthorblockA{\textit{Computer Science Department} \\
\textit{Worcester Polytechnic Institute}\\
Worcester, Massachusetts, USA \\
yli29@wpi.edu}
\and
\IEEEauthorblockN{Bohan Jiang}
\IEEEauthorblockA{\textit{Computer Science and Engineering} \\
\textit{Arizona State University}\\
Tempe, Arizona, USA \\
bjiang14@asu.edu }
\linebreakand  
% \and
\IEEEauthorblockN{Kai Shu}
\IEEEauthorblockA{\textit{Department of Computer Science} \\
\textit{Illinois Institute of Technology}\\
Chicago, Illinois, United States \\
kshu@iit.edu}
\and
\IEEEauthorblockN{Huan Liu}
\IEEEauthorblockA{\textit{Computer Science and Engineering} \\
\textit{Arizona State University}\\
Tempe, Arizona, USA \\
huanliu@asu.edu }
}
% \and
% \and

\maketitle
% \title{Covid-19 Datasets}
% \author{AAAI Press\\
% Association for the Advancement of Artificial Intelligence\\
% 2275 East Bayshore Road, Suite 160\\
% Palo Alto, California 94303\\
% }   

\maketitle
\begin{abstract}
\begin{quote}
The COVID-19 epidemic is considered as the global health crisis of the whole society and the greatest challenge mankind faced since World War Two. Unfortunately, the fake news about COVID-19 is spreading as fast as the virus itself. The incorrect health measurements, anxiety, and hate speeches will have bad consequences on people's physical health, as well as their mental health in the whole world. To help better combat the COVID-19 fake news, we propose a new fake news detection dataset {\data}\footnote{The dataset is available at https://github.com/bigheiniu/X-COVID} (Multilingual and Multidimensional COVID-19 Fake News Data Repository). This dataset provides the multilingual fake news and the relevant social context. We collect 3981 pieces of fake news content and 7192 trustworthy information from English, Spanish, Portuguese, Hindi, French and Italian, 6 different languages. We present a detailed and exploratory analysis of {\data} from different perspectives, and demonstrate the utility of {\data} in several potential application of COVID-19 fake news study on multilingual and social media.
\end{quote}
\end{abstract}

\section{Introduction}
COVID-19, an infectious disease caused by a newly discovered coronavirus\footnote{https://www.who.int/health-topics/coronavirus}, has caused more than 40 million confirmed cases and 1.2 million deaths around the world in 2020 November\footnote{https://coronavirus.1point3acres.com/}. Unfortunately, the fake news about Covid-19 has boosted the spreading of the disease and hate speech among people. For example, a couple who followed the half-backed health advice, took chloroquine phosphate to prevent COVID-19 and became ill within 20 minutes\footnote{https://www.12news.com/article/news/health/coronavirus/man-dies-after-self-medication-to-prevent-covid-19/75-3c832083-c740-41c4-9360-286391e1d095}; the racist linked the COVID-19 pandemic to Asian and people of Asian descent and the violence attacked Asian people have increased in the United States, United Kingdom, Italy, Greece, France, and Germany\footnote{https://www.hrw.org/news/2020/05/12/covid-19-fueling-anti-asian-racism-and-xenophobia-worldwide}. To stop the spreading of COVID-19 fake news, we should first address the problem of fake news detection. 

However, identifying these COVID-19 related to fake news is non-trivial. 
There are several challenges: firstly, the COVID-19 fake news is multilingual. For example, FACTCHECK.org, a fact-checking agency, found that the fake news "COVID-19 is caused bacteria, easily treated with aspirin and coagulant." is firstly seen in  Portuguese in Brazil then has the version of  English in India and American\footnote{https://www.factcheck.org/2020/05/covid-19-isn't-caused-by-bacteria/}. The current available fake news datasets and methods are mainly focused on monolingual, omit the correlation between different languages. Thus it is necessary to have a multilingual fake news dataset to utilize rich debunked fake news language to help detect fake news in poor resource language. Second, fake news content merely provides a limited signal for spotting fake news. This is because the fake news is intentionally written to mislead readers and the difficulty in correlating multilingual fake news content. Thus, we need to explore auxiliary features except for fake news content such as social engagements and user profiles on social media. For example, people who post many vaccine conspiracy theories are more likely to transmit COVID-19 fake news. Thus, it is necessary to have a comprehensive dataset that has multilingual fake news content and their related social engagements to facilitate the COVID-19 fake news detection. However, to the best of our knowledge, existing COVID-19 fake news datasets did not cover both aspects. 

% \begin{table}[]
%     \centering
%     \caption{The fake news of \textit{Coronavirus is spread through 5G} in different languages.}
%     \begin{tabularx}{\linewidth}{|c|L|}
%         \hline
%     \textbf{Language} & \textbf{fake news Content} \\
%         \hline
%     English(en) & Coronavirus is actually an expanded exosome caused by 5G. \\ 
%         \hline
%     Spanish(es) & ¿Vacunas con ARN digitalizable que la red 5G activa? \\
%         \hline
%     Portuguese(pt) &  5G transmite coronavírus\\
%         \hline
%         Hindi(hi) &  
% \begin{hindi}
%     कोरोना वायरस \end{hindi} 5G \begin{hindi} टेक्नॉलॉजी से भी संक्रमण फैल सकता है. 
%     \end{hindi} \\
%         \hline
%     Italian(it) & Coronavirus. Il 5G penetra nelle cellule indebolendo il sistema immunitario. \\
%         \hline
%     French(fr) &  Le virus a frapper plus fort là où se trouve la 5G. \\

%         \hline
%     \end{tabularx}

%     \label{tab:fake_news}
% \end{table}

Therefore, in this paper, we present a fake news dataset {\data} which contains fake news content, social engagements, and spatial-temporal information in 6 different languages. This dataset will bring several advantages to combating global COVID-19 fake news. First, the multilingual dataset provides an opportunity for cross-language fake news detection. Secondly, a rich set of features facilitate the research on multi-modal(visual and textual) fake news detection and boosting the fake news performance by including auxiliary social context. Thirdly, the temporal information provides an idea experiment data for early fake news detection. Researchers can flexibly set the cutoff time periods to test the sensitivity of the proposed model. Fourthly, researchers can investigate the fake news diffusion process on the languages and the social network for developing intervention strategies to mitigate the bad impacts of fake news~\cite{shu2018studying}.  The main contribution of this dataset are as follows:
\begin{itemize}
    \item We provide a multilingual and multidimensional fake news dataset  ~{\data} to facilitate the fake news detection and mitigation;
    \item We conduct extensive exploration analysis on~{\data} from a different perspective to demonstrate the quality of this dataset, and provide baseline methods for multilingual fake news detection, and
    \item We discuss benefits and propose insights for the fake news detection research on multilingualism and social media with {\data}. 
    
\end{itemize}

This rest of this paper is organized as follows. We review the related work in Section~\ref{sec:related_work}. The detail dataset construction and collection are presented in Section~\ref{sec:data_collection}. The exploring data analysis and fake news detection baselines are illustrated in Section~\ref{sec:data_analysis} and Section~\ref{sec:experiment} respectively. Finally, we propose insights into multilingual fake news detection in Section~\ref{sec:insight} and conclude in Section~\ref{sec:conclude}.

\section{Background and Related Work}\label{sec:related_work}
The COVID-19 fake news is a global threat now. Different languages of fake news is an explosion on social media. Most of them are intentionally written to mislead readers. To better combat the COVID-19 fake news, a multilingual and comprehensive dataset for developing fake news detection methods is necessary. Although there are many fake news datasets, most of them are either monolingual or only with linguistic features. To relieve the threat of fake news during the pandemic, we propose a dataset {\data}, which not only contains multilingual fake news, but also multi-dimensional features including news contents and social engagements. To be clarified, we list the detailed introduction of the related fake news dataset in the following. 
\begin{table*}[tbh!]
    \centering
    \caption{Comparison with existing COVID-19 fake news datasets.}
    \begin{tabular}{|c|c|c|c|c|c|c|c|c|c|}
    \hline
        \multirow{2}{*}{\backslashbox{\textbf{Dataset}}{\textbf{Features}}} &  \multicolumn{3}{c|}{\textbf{News Content}} & 
        \multicolumn{4}{c|}{\textbf{Social Context}}  & \multicolumn{2}{c|}{\textbf{Spatial-Temporal}} \\
    \cline{2-10}
     & Multilingual  & Linguistic & Visual & Tweet & Response & User & Network & Spatial & Temporal\\
     \hline
     Liar & -  & $\surd$& -  & - & - & - & - & - & - \\
    FakeNewsNet & -  &  $\surd$ &  $\surd$ & $\surd$ & $\surd$ & $\surd$ & $\surd$ & $\surd$ & $\surd$ \\

    \hline
    FakeCovid & $\surd$ & $\surd$ & - & - & - & - & -& $\surd$ & $\surd$ \\ 
    % FakeNewsNet & & $\surd$ & $\surd$ & $\surd$ & $\surd$ & $\surd$\\ 
    ReCOVery & -  & $\surd$ & $\surd$ & $\surd$ & $\surd$ & $\surd$ & $\surd$  & $\surd$ & $\surd$  \\ 
    CoAID & -  & $\surd$&  $\surd$ & $\surd$ & $\surd$ & - & - & - & $\surd$ \\
    CMU-MisCOV19 & - & $\surd$& - &  $\surd$ & - & - & - & $\surd$ & $\surd$ \\
    covid19-datasets & -  & $\surd$ & - &$\surd$ & - & - &- & - & - \\
    \hline
    {\data} & $\surd$  & $\surd$ & $\surd$ & $\surd$ & $\surd$ & $\surd$ & $\surd$ & $\surd$ & $\surd$   \\
    \hline
    \end{tabular}

    \label{tab:related_dataset}
\end{table*}

\begin{itemize}
    \item Liar\cite{wang2017liar}: There are 12.8k annotated short statements with various contexts PolitiFact. Each statement contains the statement content, speaker, context, label, and detailed justification from professional editors. 
    
    \item FakeNewsNet~\cite{shu2018fakenewsnet}: This dataset includes the fact-checking article, source article which was debunked or supported in the fact-checking website, and related social engagements. This dataset is collected from PolitiFact\footnote{www.politifact.com} and GossipCop\footnote{www.gossipcop.com} with total 23,196 news pieces and 690,732 tweets.

    \item FakeCovid~\cite{shahi2020fakecovid}: There are  5182 pieces of COVID-19 fact-checking news pieces in 40 languages from 105 countries. It get the labeled content from  Snopes\footnote{{www.snopes.com/}} and Poynter\footnote{{www.poynter.org}}.
 
    \item ReCOVery~\cite{zhou2020recovery}: This dataset is used for news credibility classification. It collects the incredible news from the domain listed in NewsGaurd\footnote{{www.newsguardtech.com}}. This dataset contains the news content and related social context in Twitter. There are 2, 029 news pieces and 140, 820 tweets in this dataset. 
    \item CoAID~\cite{cui2020coaid}: This dataset contains the labeled news article, short claim, social post and the related social engagements. There are 1, 896 news pieces, 516 social platform posts and 183, 569 related user engagements.
    %  \item Covid-19~\cite{chen2020covid}: This is the first public coronavirus twitter dataset which contains 123,113,914 multilingual tweets. However, they only collected raw data from twitter without labels.
    %  \item GeoCoV19~\cite{qazi2020geocov19}: This dataset contains over 524 million multilingual tweets, with over 94\% tweets have geolocation information. They do not provide labeled news and social contents.
    % \item NAIST COVID~\cite{gao2020naist}: This dataset contains 16,250,038 tweets in English, 9,501,866 tweets in Japanese and 173,869 Weibo posts in Chinese. It is a spatial covid-19 dataset collected from social media platforms without labels. 
    \item CMU-MisCOV19~\cite{memon2020characterizing}: This is a covid-19 related dataset with 4,573 annotated tweets in English. They classify the users into informed, misinformed and irrelevant groups.
     \item covid19-datasets~\cite{inuwa2020curated}: The authors utilize the COVID-19 myth related keywords to collect the fake tweets.

\end{itemize}

From Table~\ref{tab:related_dataset}, we can find that no existing fake news datasets can afford the multilingual fake news and comprehensive news content and social engagements. There are still some limitations to the existing datasets that we want to address in our proposed dataset. For example, FakeCovid labeled news pieces into fake and not fake which contains partly fake, half true, missing evidence, and so on. The news contents in FakeNewsNet contains noise since some of them are collected from Google Search result which often mentions similar but unrelated news pieces. ReCOVery labels each news piece as credible and incredibly based on the news source, rather than the human experts separately label each news pieces. CoAID mostly keeps the title of the fake news and much fake news misses the social engagements. 

To address the aforementioned limitations of the existing datasets, we provide a new multilingual and multi-dimensional dataset {\data} which covers 6 languages and contains the information from the fake news content to the related social engagements.

\section{Data Collection}\label{sec:data_collection}
\begin{figure*}[t!]
    \centering
    \includegraphics[width=0.9\linewidth]{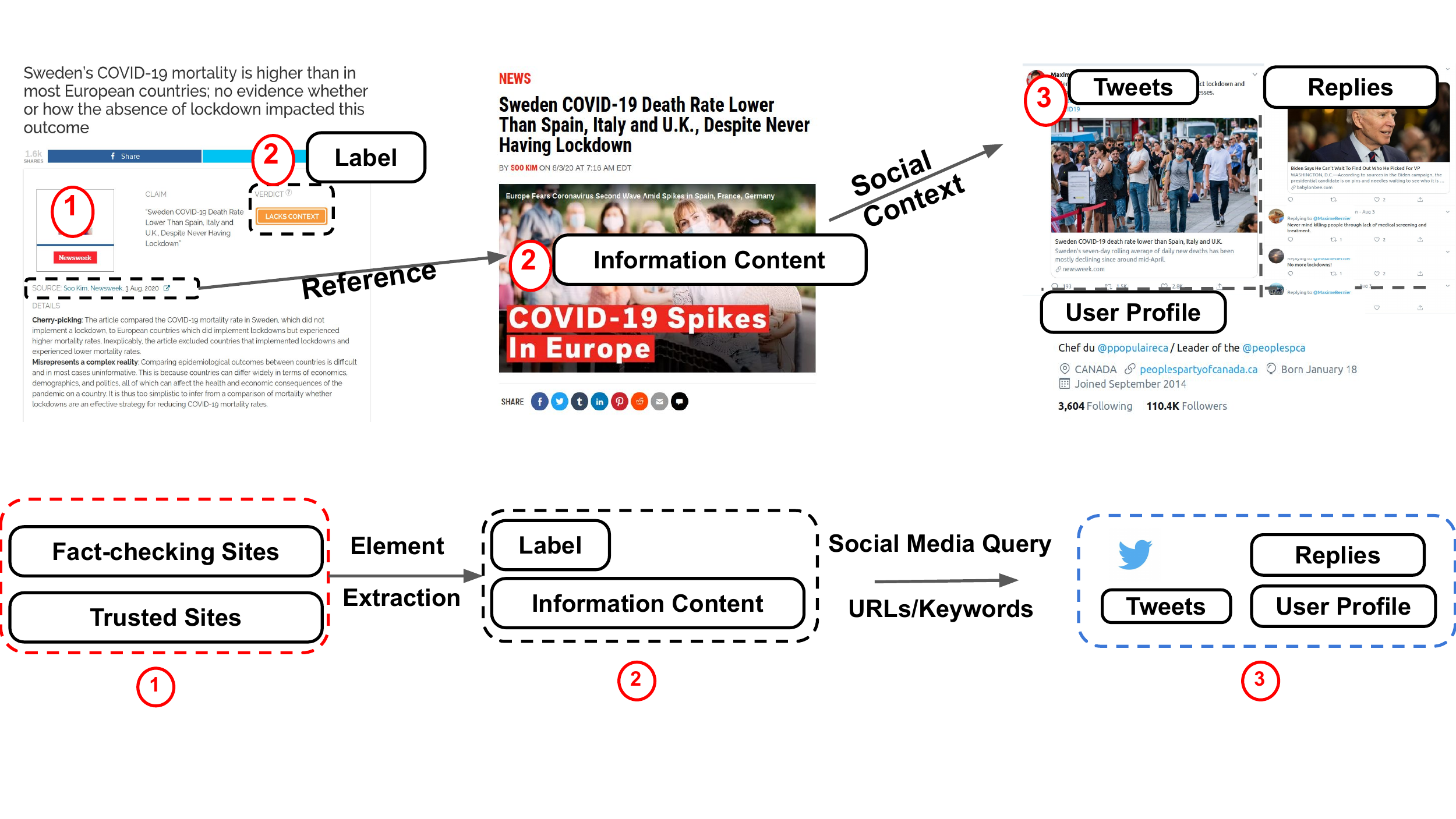}
    \caption{The data collection process for {\data} with screenshot example. }
    \label{fig:collect_pipeline}
\end{figure*}

In this section, we introduce the whole procedure of data collection, including fake news content and social context. The whole process is depicted in Figure~\ref{fig:collect_pipeline}.
% All the fake news and real news content contain the 2-3 sentences short claims and long news article. 

\subsection{News Content Collection}
% As shown in Figure~\ref{fig:pipeline}, the fake news collection includes the ground truths and source news content. 
As shown in Figure~\ref{fig:collect_pipeline}, we need to firstly get the reliable labels from the fact-checking websites, and then retrieve the source content from these websites. We collect the veracity labels from Snopes\footnote{{www.snopes.com}} and Poynter\footnote{{www.poynter.org/coronavirusfactsalliance/}} where the domain expert and journalists review the information and provide the fact-checking evaluation results as fake or real. Snopes is an independent publication owned by Snopes Media Group and mainly contains English content. 

Poynter is an international fact-checking network~(IFCN) alliance unifying 96 different fact-checking agencies like PolitiFact\footnote{{www.politifact.com}}, FullFact\footnote{{fullfact.org/}} and etc, in 40 languages. 

To keep the quantity of each language, we only filter languages like English~(en), Spanish~(es), Portuguese~(pt), Hindi~(hi), French~(fr), and Italian~(it).  
Because the Poynter website only displays the translated English claims, we set the language for each claim based on the language used in the fact-checking article. 
After collecting the reliable label, we set heuristic crawling strategies for each fact-checking website to fetch the source content URL from the fact-checking websites. 
In some cases, the source content URL may be no longer available. To resolve the problem, we check the archived website\footnote{{archive.is}} to see whether the page is archived or not. If not, we will consider the claim as the content of fake news. 
% Directly referencing the source content based on the URLs listed in the fact-checking articles instead of utilizing the Google Search result, and removing nonexistent URLs after checking the archive website greatly reduce the noise among different languages. 
% From our initial observation, we find xx\% of the label are fake. This is because the purpose of Poynter is to debunk fake news. 

Since most news pieces in Poynter and Snopes are fake news, to balance the dataset for each language, we choose several official health websites and collect the COVID-19 related news in these websites as additional real information. To filter unrelated information, we collect the news piece whose title contains any of the keywords \textit{COVID-19}, \textit{Coronavrius} and \textit{SARS-CoV-2}. The reliable websites for each language are listed in Appendix Table~\ref{tab:detail_news}. After we get the source URLs, we utilize the Newspaper3k\footnote{https://newspaper.readthedocs.io/en/latest/} to crawl the content and its meta-information. 

\begin{table}[]
    \centering
    \caption{Description of the features including in the dataset}
    \begin{tabular}{|c|j|}
        \hline
         \textbf{Category} &  \textbf{Features} \\
         \hline
         Fact-checking Reviews & Fact-checking URL, Veracity Label, Debunked Explanations, Twitter Search Query  \\
         \hline
         Source Contents & URL, Language, Location, Release Date, Text Content, Image \\
         \hline
         Social Engagements & Tweets, Replies, Retweets \\
         \hline 
         Twitter Users & Profiles, Timelines, Location, Followers, Friends
         \\
                  \hline
         
    \end{tabular}
    
    \label{tab:my_label}
\end{table}
% \begin{itemize}
%     \item \textbf{text}: the main text body of the information. 
%     \item \textbf{publish date}: the date when the information is published online. 
%     \item \textbf{publish location}: the location where the information is published online. 
%     \item \textbf{authors}: the name of the authors who wrote this information. If the information is social media posts, like tweets or Facebook posts, we use their screen name as the unique identifier; if it is the news piece, we use the authors name externally provide by the news website. 
%     \item \textbf{Agency}: the site or domain where the information is posted.
% \end{itemize}

It should be noticed that the source of both fake news and real news include social media posts like Facebook, Twitter, Instagram, WhatsApp, etc, and news article posted in blogger and traditional news agencies. 

\subsection{User Social Engagement}
    As shown in Figure~\ref{fig:collect_pipeline}, we collect the user social engagements from the social platform based on the news content. Specifically, we form the search query based on the URL, the headline and the first sentence of the source content then use the Twitter advanced search API\footnote{https://twitter.com/search-advanced?lang=en} through twarc~\footnote{https://github.com/DocNow/twarc} to collect the user social engagements.  To reduce the search noise, we remove the special character, negative word, utilize the TFIDF~\cite{rajaraman_ullman_2011} to extract the important words, and lastly check the query manually.
    The social engagements include the tweets which directly mention the news pieces, and the replies and retweets responding to these tweets. After we obtain the related tweets from the advanced search result, we collect the tweets' replies and retweets. Due to the fact that Twitter's API does not support getting replies, we approximately utilize the tweet's ID as the search query, which can only obtain the replies sent in the last week\footnote{https://github.com/DocNow/twarc}. In the end, we fetch all users' profiles, network connection, and the timeline of who engages in the news dissemination process.

% \section{Data Structure}
% In this section, we clarify the attributes of fake news content and its related social context. 
% \subsection{fake news Content}
% To differentiate each news pieces, we auto generate a distinct fake news content ID for each fake news. We 

\section{Data Analysis}\label{sec:data_analysis}
In this section, we will demonstrate the quality of the {\data} through statistical analysis and visualization. Because {\data} contains multi-dimensional information which can be used as features to identify the fake news, we separately make comparison among real news and fake news in source content, social context, and language spatial-temporal information. We also select several fake news detection methods as baseline methods for further research. The detailed statistical information of our dataset is demonstrated in Table~\ref{tab:data_statistical}. 

\begin{table*}[!ht]
    \centering
    \caption{Statistics of {\data}}
\scalebox{0.8}{\begin{tabular}{|l|c|c|c|c|c|c|c|c|c|c|c|c|}
    \thickhline
\multirow{2}{*}{Category} & \multicolumn{6}{c|}{\centering{\textbf{Fake}}}   & \multicolumn{6}{c|}{\centering{\textbf{Real}}}  \\
% English, Spanish, Portuguese, Hindi, French, Turkish and Italian
\cline{2-7} \cline{8-13}  & 
\textbf{en} & \textbf{es} & \textbf{pt} & \textbf{hi} & \textbf{fr} & \textbf{it}  &  \textbf{en} & \textbf{es} & \textbf{pt} & \textbf{hi} & \textbf{fr} & \textbf{it}   \\ 
\hline
\# Source Content & 2,168 & 808 & 371 & 336 & 189 & 109  & 2,114 & 2,405 & 713 & 1,023 & 392 &  937   \\
\thickhline
\# Tweets & 32,811 & 21,911 & 15,738 & 1,143 & 2,821 & 750 & 26,565 &	1,553 &	268 &	1,205 & 166 & 369  \\
\hline
\# Replies & 25,888 & 15,222 & 14,679 & 1,015 & 4,459 & 1,323
&  18,749 &	33,939 & 858 & 11,381 & 5,095 & 7,816  \\
\hline
\# Retweets & 43,048 &	32,986 & 20,377 & 1,677 & 6,552 & 1,323 &  41,270 & 74,511 & 2,393 & 42,477 & 5,565 & 17,599 \\
\hline
% Retweets & x & x & x & x & x & x & x & & x & x & x & x & x & x & x \\
\# Twitter Users & 37,148 & 24,644 & 14,691  & 1,536 & 4,760 & 978 & 19,225 & 4,180 & 203 & 1,972 & 86 & 1,291  \\
\thickhline
    \end{tabular}}
    \label{tab:data_statistical}
\end{table*}

\subsection{Source Content Analysis}
Since the malicious users mostly manipulate the text content to mislead the audience, there stay text clues in the fake news content. We reveal these clues through the word cloud and the visualization of semantic representation and make a comparison among the fake news and real news.

\begin{figure*}[tbh!]
    \centering
%  \subfigure[\textbf{All language Fake}]{
% 	\includegraphics[width=0.22\linewidth]{figure/all_word_cloud_fake.png}}
\subfigure[en Fake]{
\includegraphics[width=.14\linewidth]{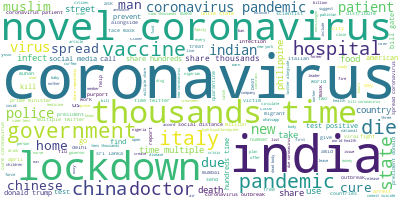}
% \label{fig:all_fake}
}
\subfigure[es Fake]{
\includegraphics[width=.14\linewidth]{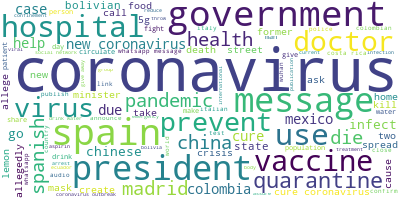}
}
\subfigure[pt Fake]{
\includegraphics[width=.14\linewidth]{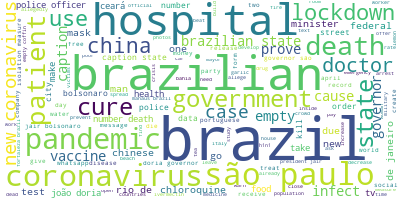}
}
\subfigure[hi Fake]{
\includegraphics[width=.14\linewidth]{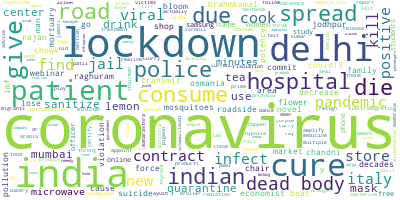}
}
\subfigure[fr Fake]{
\includegraphics[width=.14\linewidth]{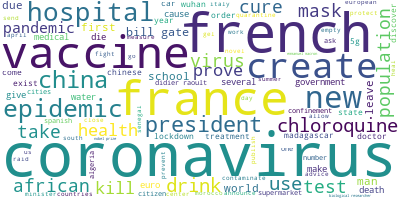}
}
\subfigure[it Fake]{
\includegraphics[width=.14\linewidth]{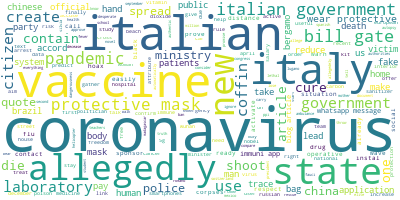}
}

% \subfigure[\textbf{All language Real}]{
% 	\includegraphics[width=0.22\linewidth]{figure/text_word_cloud_real.pdf}}
	
\subfigure[en Real]{
\includegraphics[width=.14\linewidth]{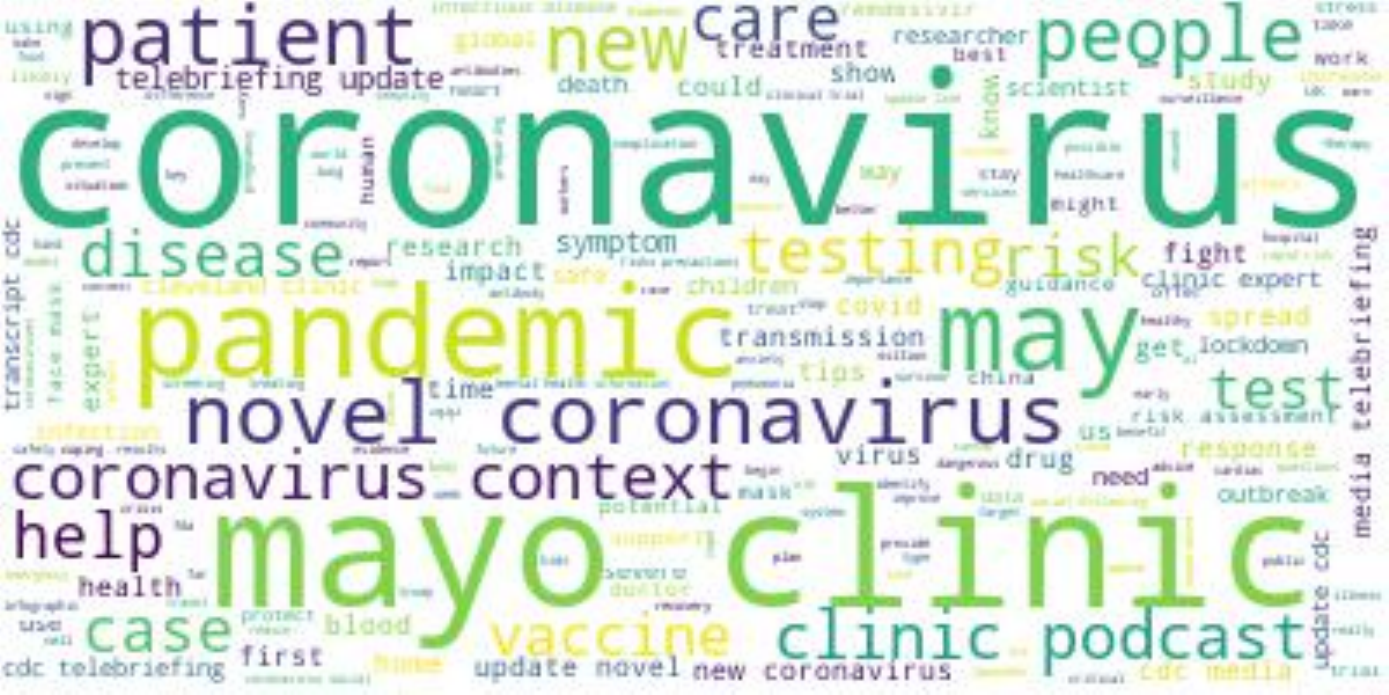}
}
\subfigure[es Real]{
\includegraphics[width=.14\linewidth]{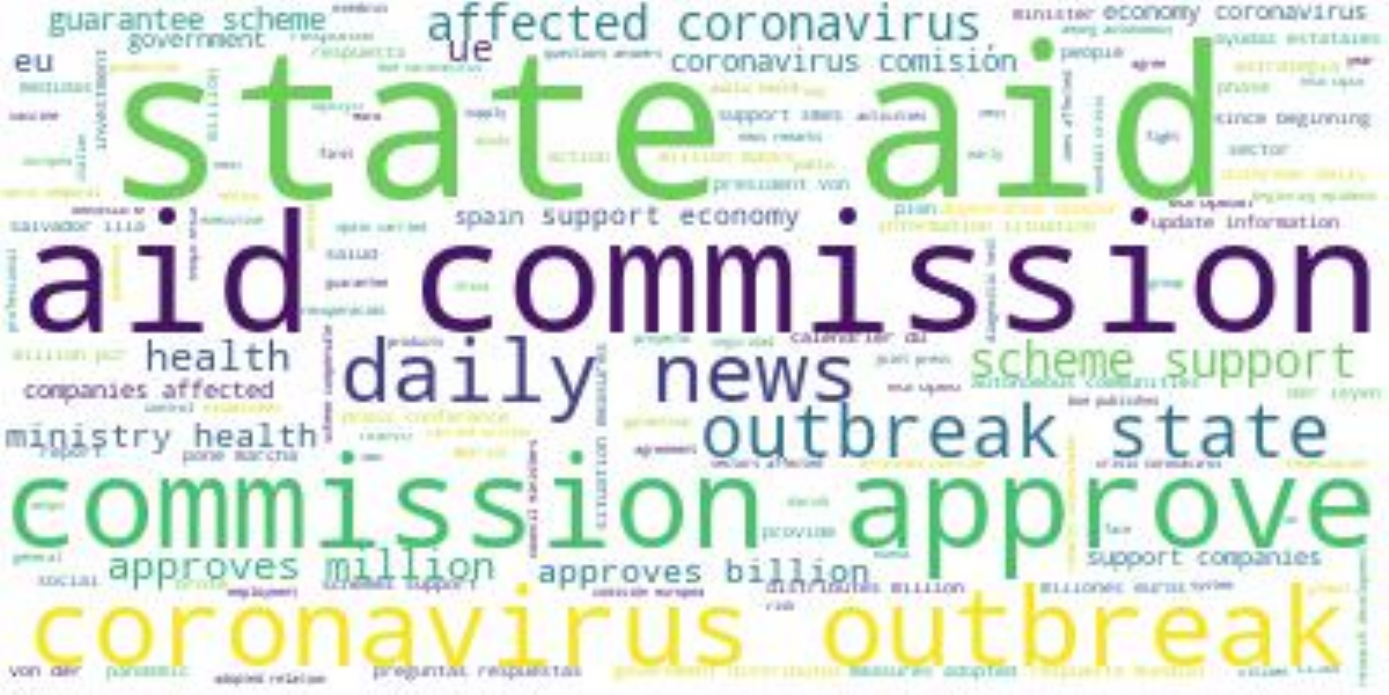}
}
\subfigure[pt Real]{
\includegraphics[width=.14\linewidth]{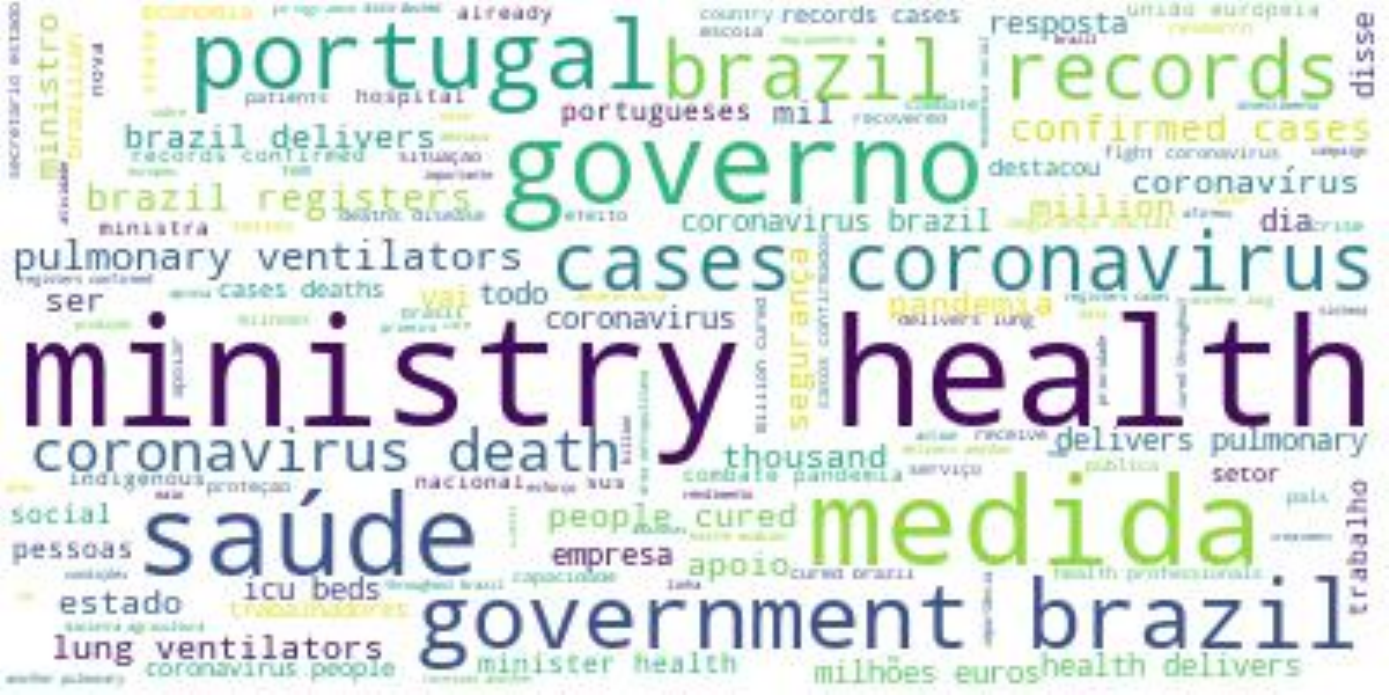}
}
\subfigure[hi Real]{
\includegraphics[width=.14\linewidth]{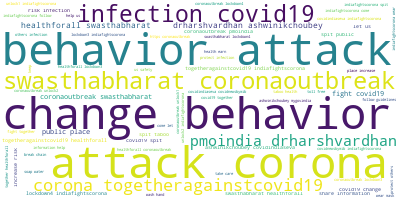}
}
\subfigure[fr Real]{
\includegraphics[width=.14\linewidth]{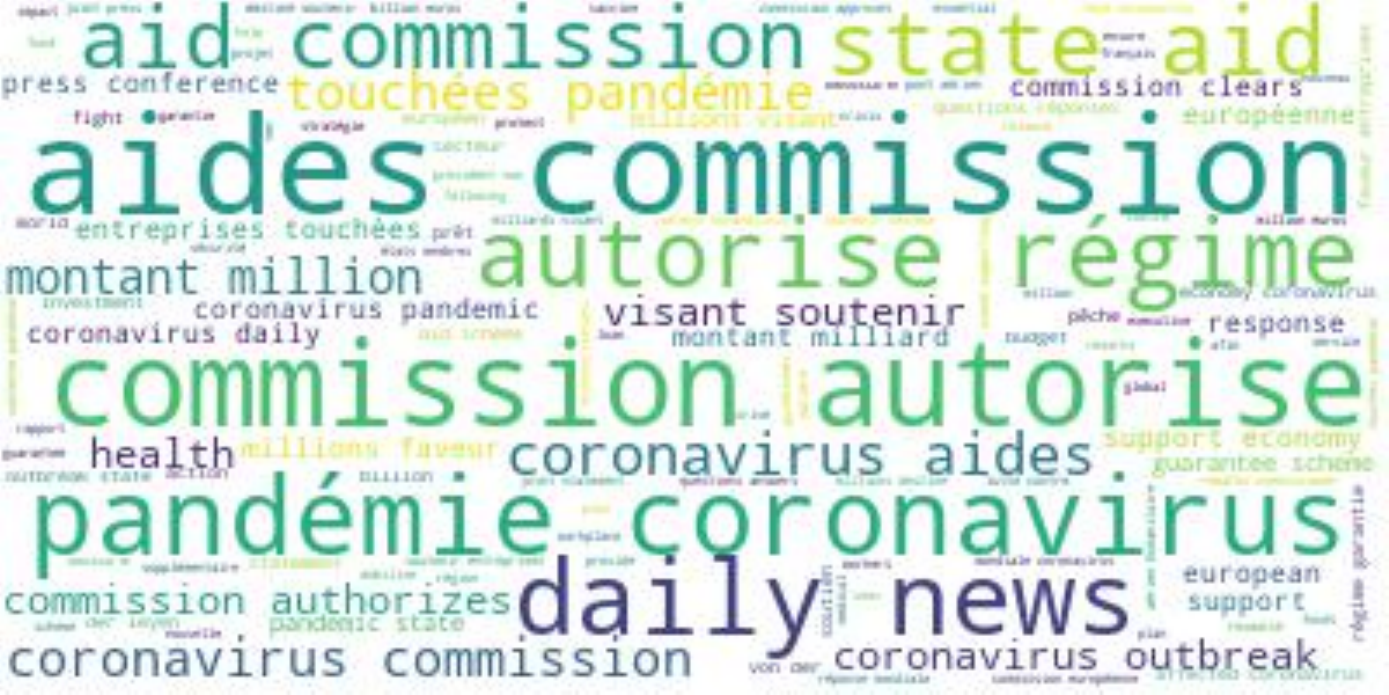}
}
\subfigure[it Real]{
\includegraphics[width=.14\linewidth]{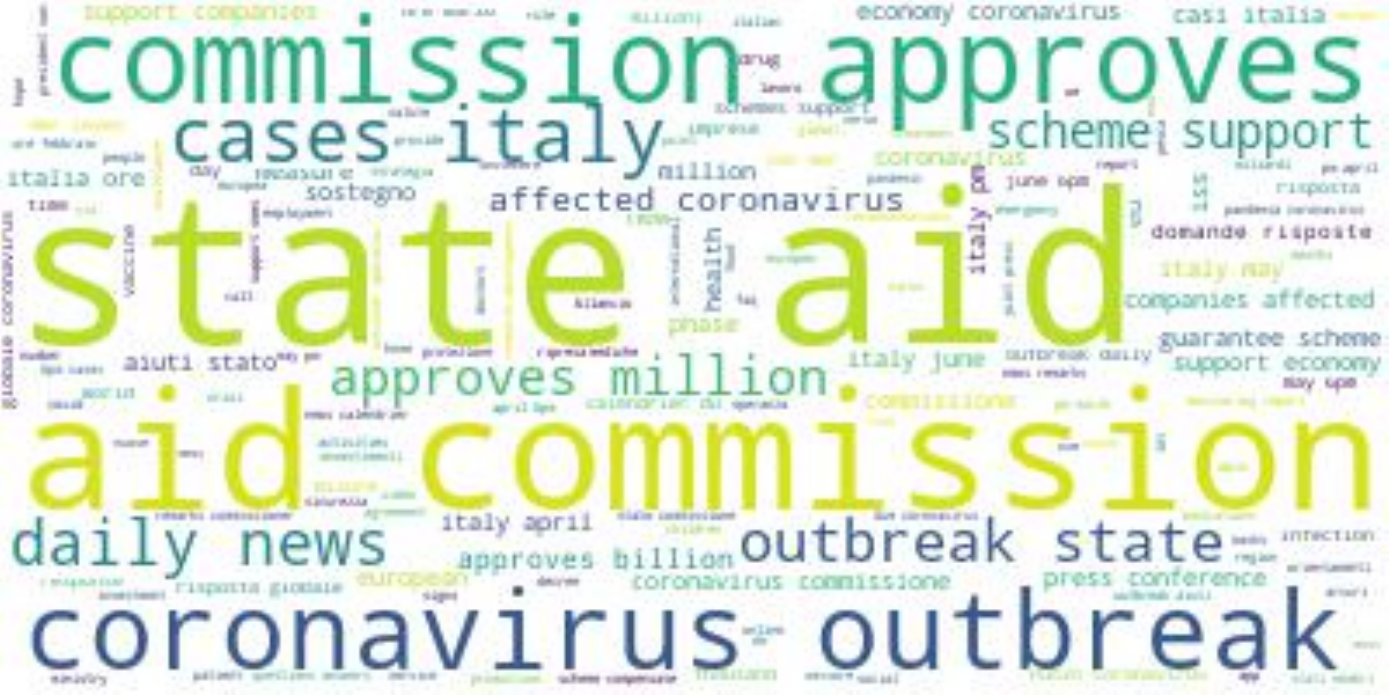}
}

    \caption{Word Cloud of the fake news and real news in different languages. All the tokens are translated into English. }
    \label{fig:word_cloud}
\end{figure*}

% We use latent Dirichlet allocation (LDA) in Gensim\footnote{https://radimrehurek.com/gensim/} to extract 10 topics for each language separately. The topic distribution visualization and the topic keyword are shown in Figure~\ref{fig:topic_distri} and Table~\ref{tab:topic}, respectively. We can find out that xxx. 
In Figure~\ref{fig:word_cloud}, we visualize the most frequent words for each language. Non-English languages are translated into English for comparison. From Figure~\ref{fig:word_cloud}, we can find the fake news often mentions the medical-related words like \textit{doctor}, \textit{hospital} and \textit{vaccine}  across languages. This is because these places are the front line of defending Coronavirus, malicious users will transmit this fake news to spread fear and anxiety.  The fake news also mentions the country name like \textit{India}, \textit{China}, \textit{Spain}, \textit{Brazil} and et al. While, the real news often mentions the keywords like \textit{test} and \textit{patient}. Besides, we also observe the topic similarity and difference among languages. For example,  languages like ``es", ``fr", and ``it", they all talk about the welfare like \textit{commission} and \textit{aid} while other languages do not mentions these phrases. Although there is a topic difference between the fake news and real news, it is not consistent across languages and meanwhile, it cannot be directly applied to a single piece of text\cite{shu2017fake}. Thus it is necessary to learn a better representation of these contents and include auxiliary features into detection like the social context. 
 
\begin{figure}[tbh!]
    \centering
    \includegraphics[width=1\linewidth]{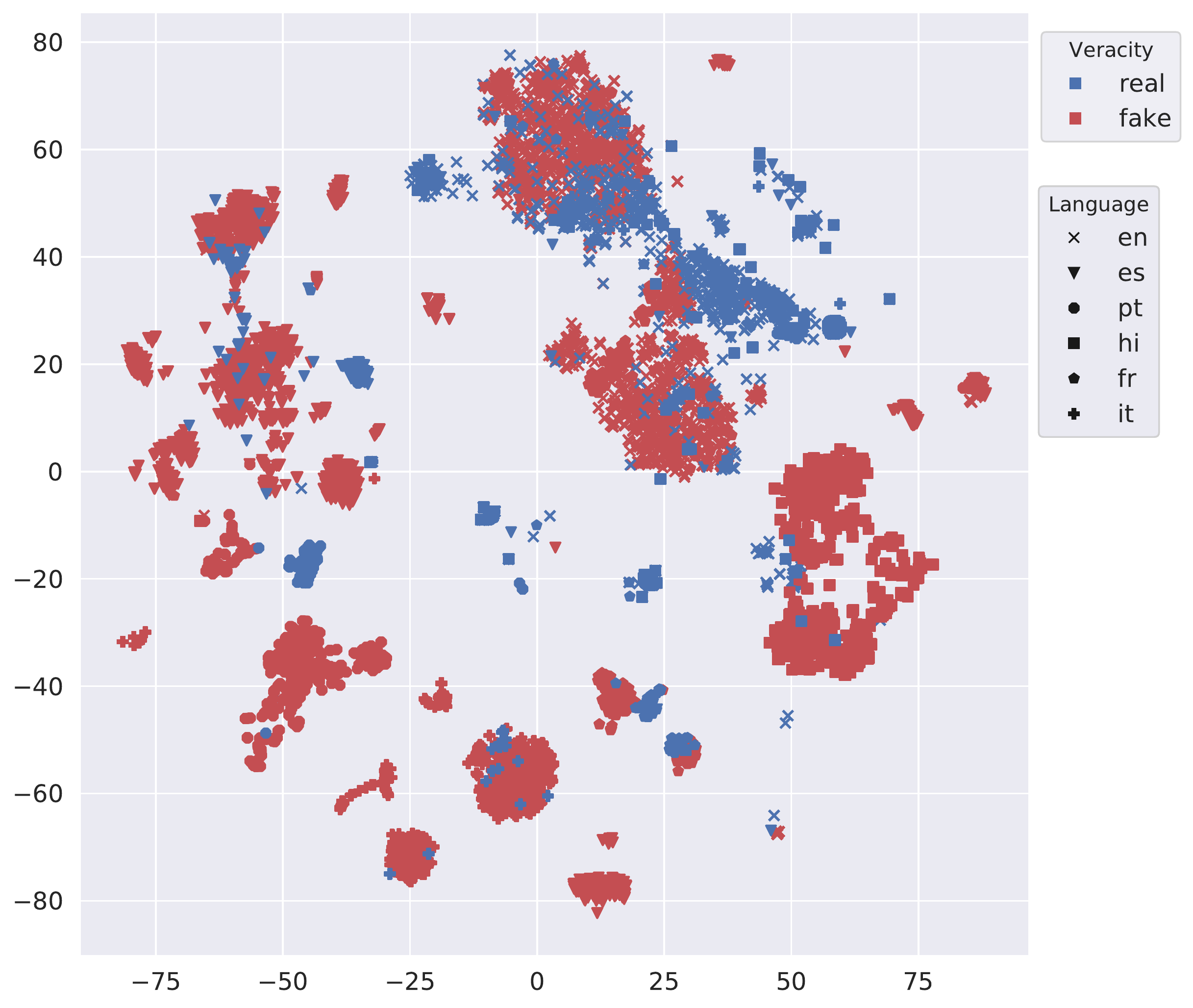}
    \caption{The visualization of the semantic representation of the fake news and real news.}
    \label{fig:semantic_vis}
\end{figure}

 Also, to understand the semantic representation difference between the fake news and real news, we visualize the hidden representation of these contents in Figure~\ref{fig:semantic_vis}. 
We firstly utilize multi-lingual RoBERTa\footnote{https://huggingface.co/xlm-roberta-base} to learn the representation of the content and utilize the t-SNE~\cite{vanDerMaaten2008} to visualize these hidden representations. From Figure~\ref{fig:semantic_vis}, we can find that there are some spreadable fake news and real news clusters, and the center upper right corner is mixed with these two labels. This observation indicates the necessity for better feature representation across languages and the difficulty in detecting fake news only on the content. 

\subsection{Language Spatiotemporal Information}
\begin{figure*}[!tbh]
    \centering
    \includegraphics[width=0.9\linewidth]{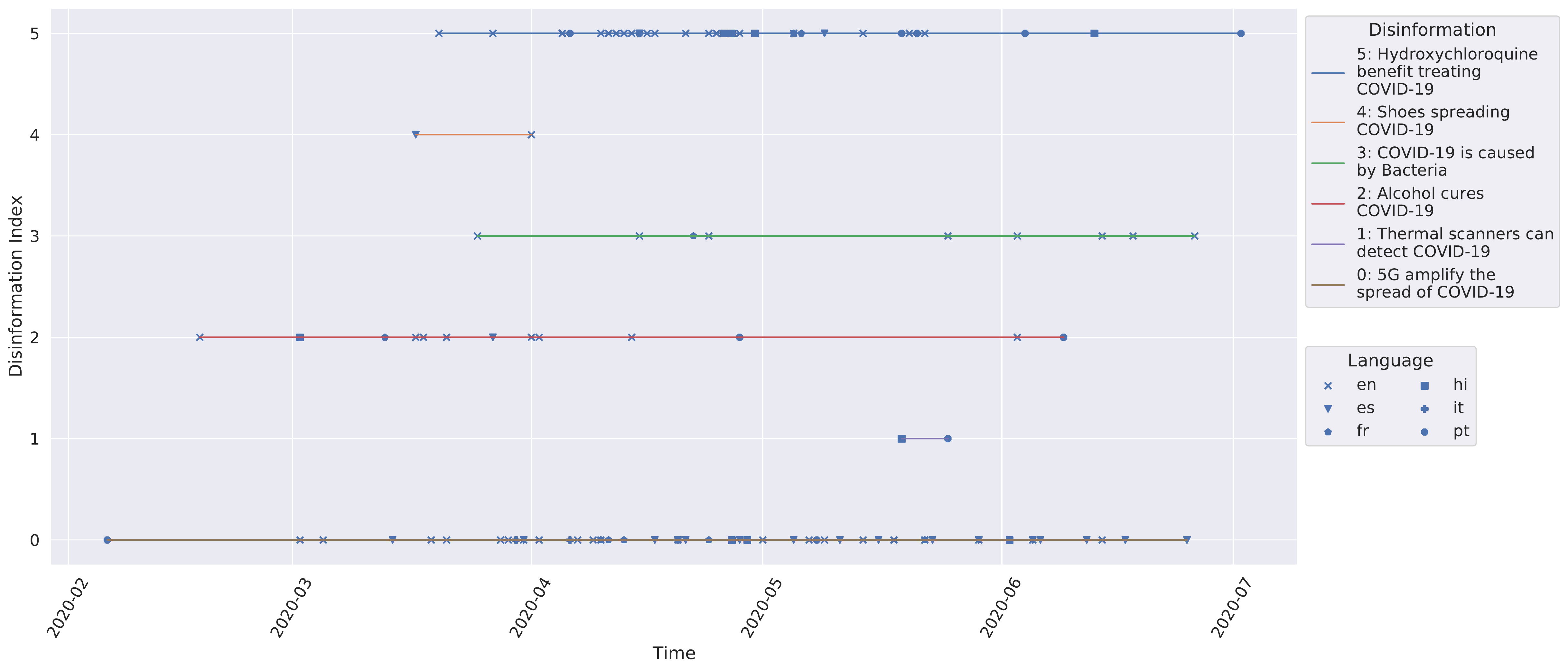}
    \caption{The fake news debunked timeline cross languages. }
    \label{fig:debunk_process}
\end{figure*}

To understand how the fake news is spread and debunked in different languages, we reveal the common fake news\footnote{https://www.who.int/emergencies/diseases/novel-coronavirus-2019/advice-for-public/myth-busters} originated and debunked timeline in Figure~\ref{fig:debunk_process}. We can find these selected fake news have been spread in different languages and there is postpone among the spreading. For example, the fake news ``\textit{Alcohol cures COVID-19}" takes about half a month to transit from English to Hindi. In addition, much fake news has many similar versions in the same language. For example, fake news like ``\textit{Hydroxychloroquine benefit treating COVID-19}" has many versions in English. This indicates the possibility of early detection cross-language and in language based on historical data.

\subsection{Measuring Social Context}
Since the social media platform provides direct access to a large amount of information which may contain the COVID-19 related fake news, the propagation networks, transition paths, and the interacted user nodes in the path. They all can provide auxiliary and language invariant information for fake news detection. 
% Since most fake news are proliferation through the social media, the social context can provide auxiliary signals to help defend the fake news. 
The monolingual social context integrated fake news models like dEFEND\cite{defend} and TCNN-URG\cite{TCNN-URG} have achieved considerable performance improvement compared with only relying on the fake news content. 
Our dataset contains three different kinds of social context: user profiles, tweet posts, and social network structure. These can provide the opportunity to explore these findings across languages. 
In the following sections, we will explore the characteristics of these features and discuss the potential utilization of fake news detection. 

\noindent\textbf{User Profiles}
The existing research\cite{shu2019role} has proven the correlation between user-profiles and fake news detection.  For example, users who are uncredible and bots-like are likely to transmit the fake news\cite{shu2020leveraging}\cite{shu2017fake}\cite{bias} and social bots play a disproportionate role in spreading fake news~\cite{Shao2018}. In this part, we will illustrate several useful features.

% Firstly, we reveal the distribution of users' credibility score in Figure~\ref{fig:credibility}. The calculation of credibility score is adopted from~\cite{credibility}. It hypothesize that non-credible users  are  more  likely  to  coordinate  with each other and form big clusters, whereas credible users are likely to form small clusters. We utilize hierarchical clustering to cluster the users based on their meta-information and take the reciprocal of the cluster size as the credibility score. From Figure~\ref{fig:credibility}, we can find users who respond to the fake news have lower credibility score compared with users who transmit the real news. 

\begin{figure}[tbh!]
    \centering
    \includegraphics[width=\linewidth]{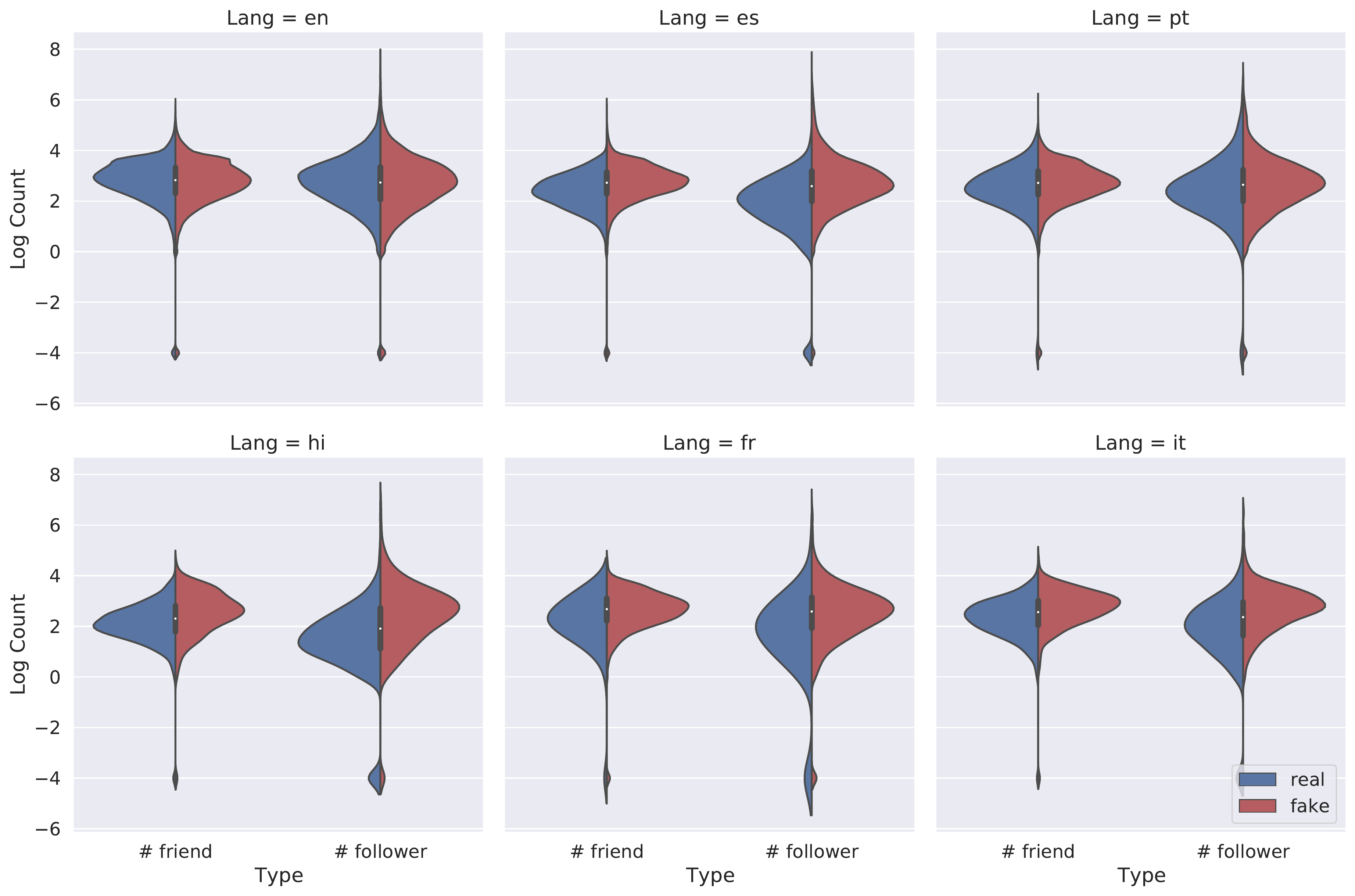}
    \caption{The $log$ count of users' \#friend and \#follower.}
    \label{fig:user_profile}
\end{figure}

% Then, we exploit users bias score in Figure~\ref{fig:bias}. The bias score is to exploit the users interests over her historical tweets\cite{bias}. The hypotheses of user bias  is that users who are left-leaning or right-leaning are more likely to share the similar interests with people hold the same bias. From Figure~\ref{fig:bias}, we observe that fake news involved users have high absolute bias scores. The above observation of fake news involved users that have low credibility score and high bias score actually meet the past research~\cite{bias},~\cite{shu2017fake} and~\cite{shu2020leveraging}. 
Firstly, we explore the social network of the users and to see whether there is a difference between the users who engage in fake news and real information. We visualize the follower and friends count of all the users in the fake news and real information in Figure~\ref{fig:user_profile}. From this figure, we can observe that users who interact with es, pt, hi, fr, and its fake news have a larger number of friends and follower than the real news with the p-value < 0.05 under statistical t-test. However, in en, there is no significant difference in the followers and friends. 

\begin{figure}
    \centering
    \includegraphics[width=\linewidth]{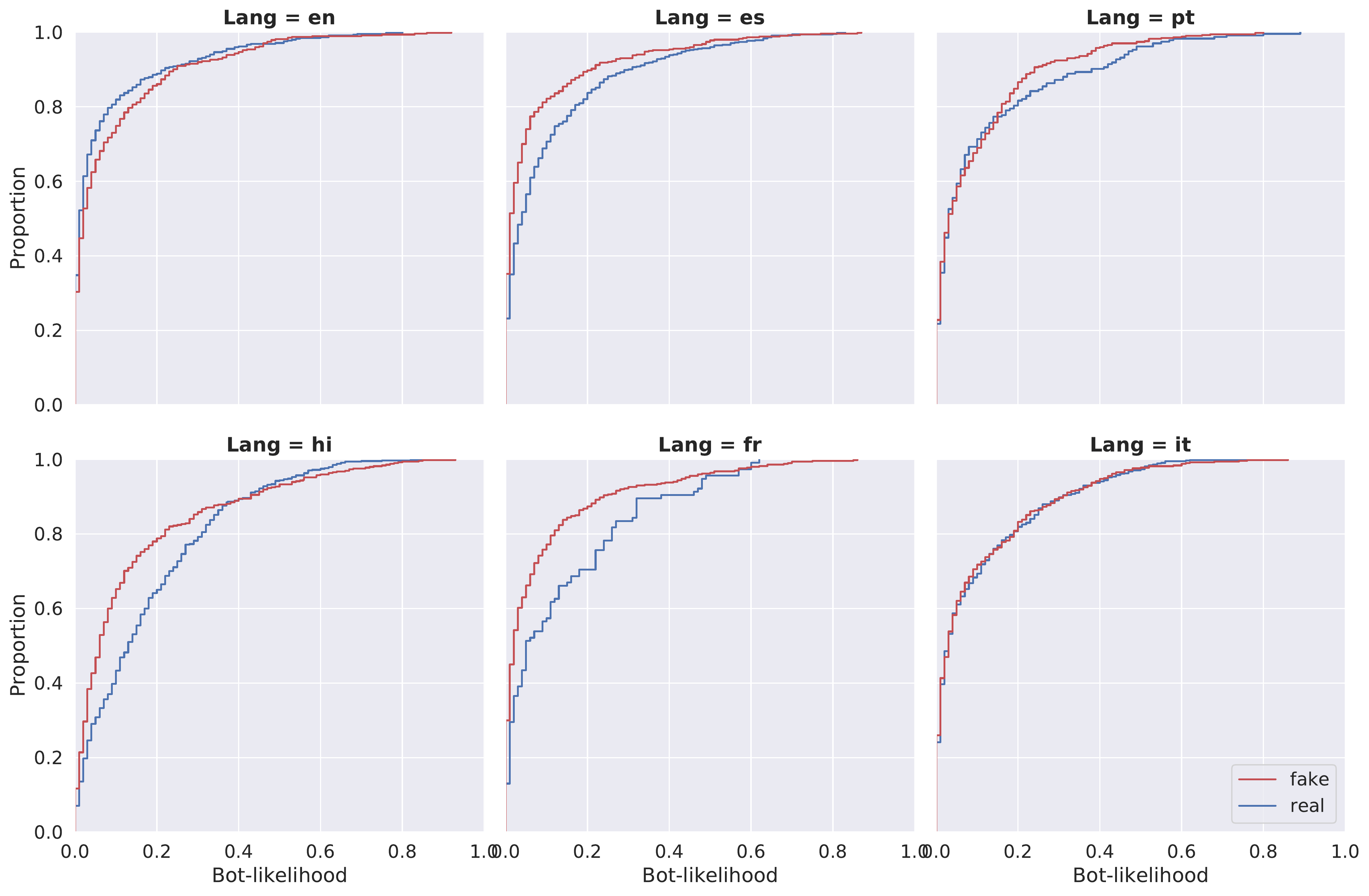}
    \caption{The cumulative bot-likelihood distribution for users engaged with fake news and real news in all languages.}
    \label{fig:bot_distribution}
\end{figure}

Lastly, we include more user profile information and to understand the bot-like probability of users in different veracity of information. For each language, we randomly sample 500 users who only respond to the fake news and another 500 users related to real news for the bot detection. For a language that contains less than 500 users, like pt, fr in real news, we take all the users in these languages. We utilize the state-of-the-art bot detection method Botometer~\cite{Davis_2016} to identify the probability of users being social bots. Botometer makes the prediction based on users' public profile, timeline, and mentions. From the cumulative distributions listed in  Figure~\ref{fig:bot_distribution}, we can find that the users who engage in fake news are slightly more likely to be bots. In languages like hi, fr, the users who have extremely large bot-likelihood (> 0.6) are more likely to interact with the fake news. This observation is also consistent with past fake news research in~\cite{shu2018fakenewsnet,dai2020ginger}. However, we also observe that bot-likelihood does not indicate the veracity of the news. For example, in es and pt, we have the opposite observation, and in it, there is no significant difference between the real news and fake news.

% Botometer; Social Bias; Credibility

\begin{figure}
    \centering
    
    \subfigure[fake news Tweets]{
    \includegraphics[width=0.43\linewidth]{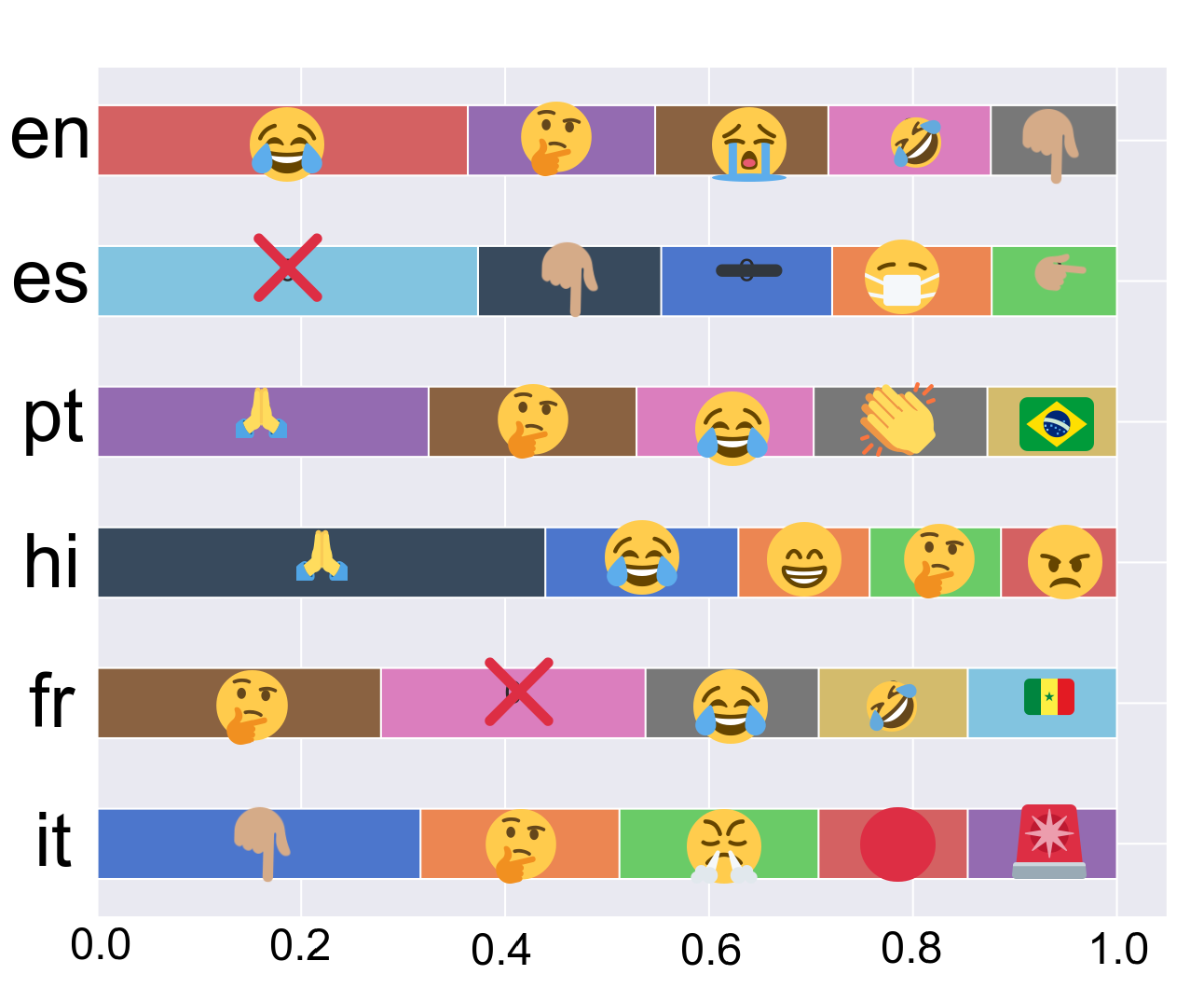}
    }
    \subfigure[Real News Tweets]{
    \includegraphics[width=0.47\linewidth]{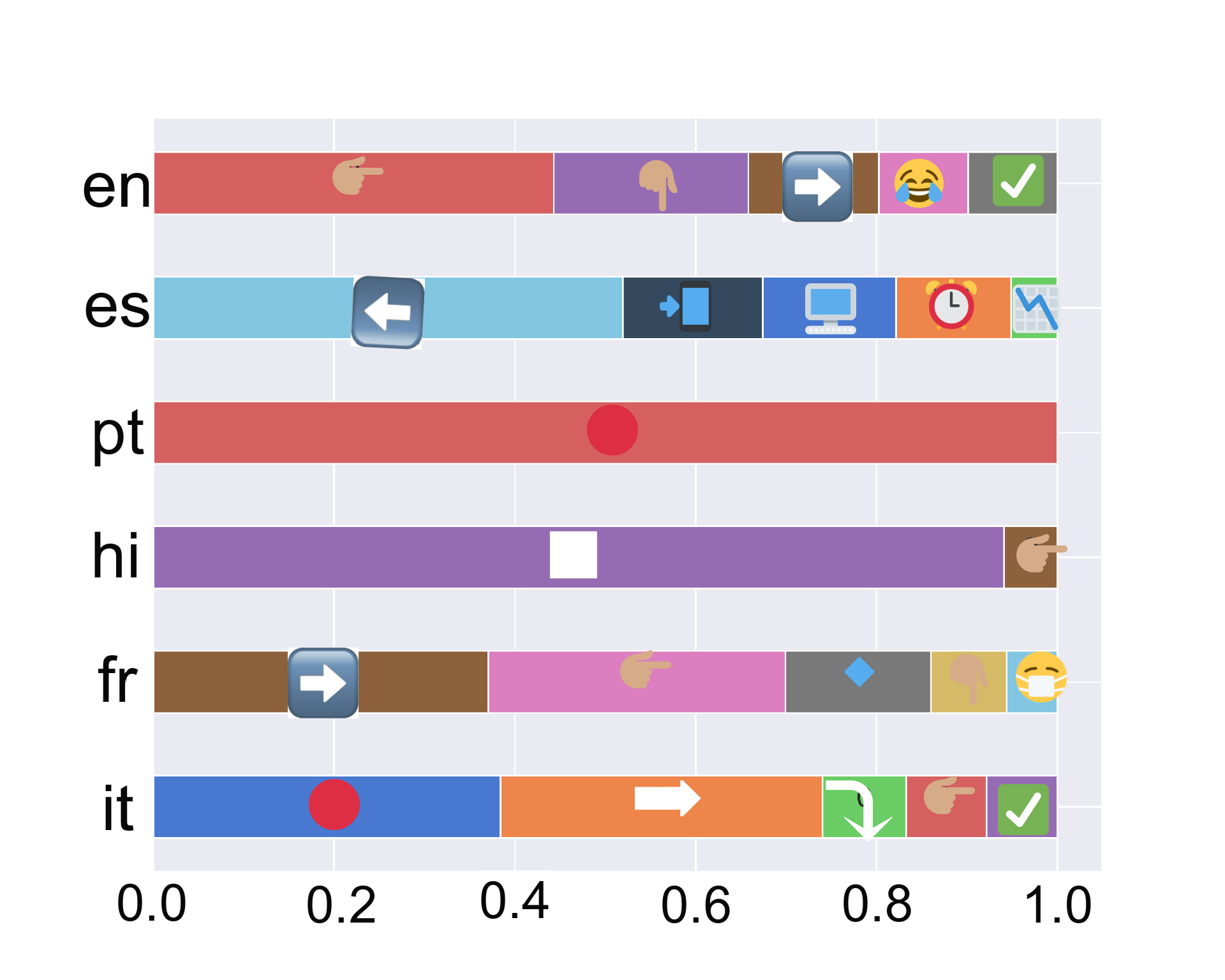}
    }
    
    \caption{Emoji distribution for  tweets in different languages.}
    \label{fig:emoji_distribution}
\end{figure}

\noindent\textbf{Tweet and Response}
In social media, people will express their emotions and focus on an event through tweets and their responses. These features can benefit the detection of fake news in general~\cite{Jin2016contradiction}\cite{qazvinian-etal-2011-rumor}.
We firstly perform the sentiment analysis on the tweets. Since there is no sentiment classification method cover these 6 languages and emoji is the proxy of the sentiment in the tweets, we reveal the distribution of emojis for tweets among different languages in Figure~\ref{fig:emoji_distribution}. Looking at the emoji of the reply tweets~(Figure~\ref{fig:emoji_distribution}), we observe that there are more emotional emoji in the tweets, like laughing in \textbf{en}, \textbf{pt}, \textbf{hi} and \textbf{fr},  and angry in \textbf{hi} and \textbf{it}. However, in the real news, the direction and enumeration emoji dominate in all languages. These observations indicate that emoji or users' emotions can benefit from fake news detection. 

\begin{figure*}[!tbh]
    \centering
    % \subfigure[\textbf{All}]{
    % \includegraphics[width=0.13\textwidth]{figure/all_tweets_reply_count.pdf}
    % }
    \subfigure[\textbf{en}]{
    \includegraphics[width=0.15\linewidth]{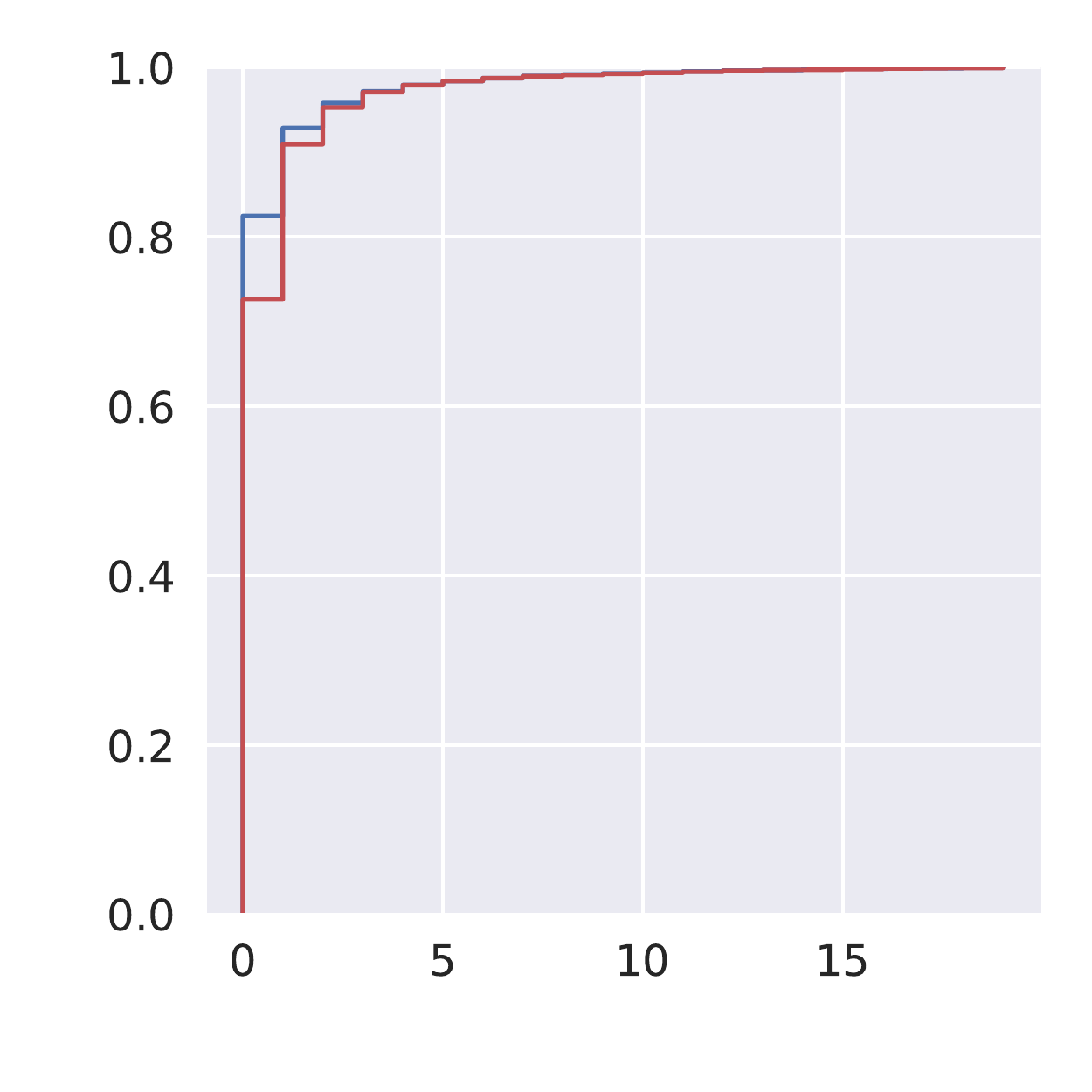}
    }
    \subfigure[\textbf{es}]{
    \includegraphics[width=0.145\linewidth]{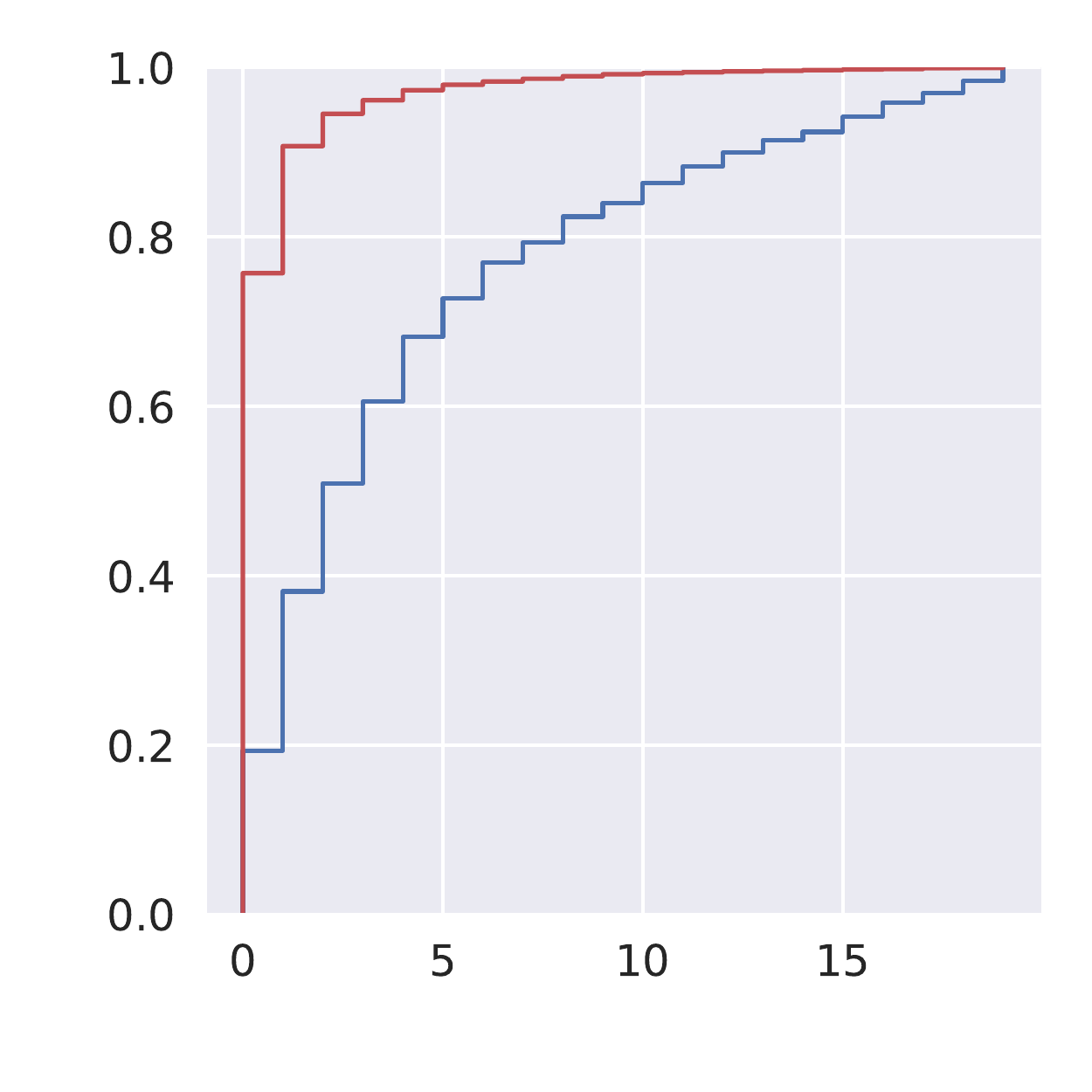}
    }
    \subfigure[\textbf{pt}]{
    \includegraphics[width=0.145\linewidth]{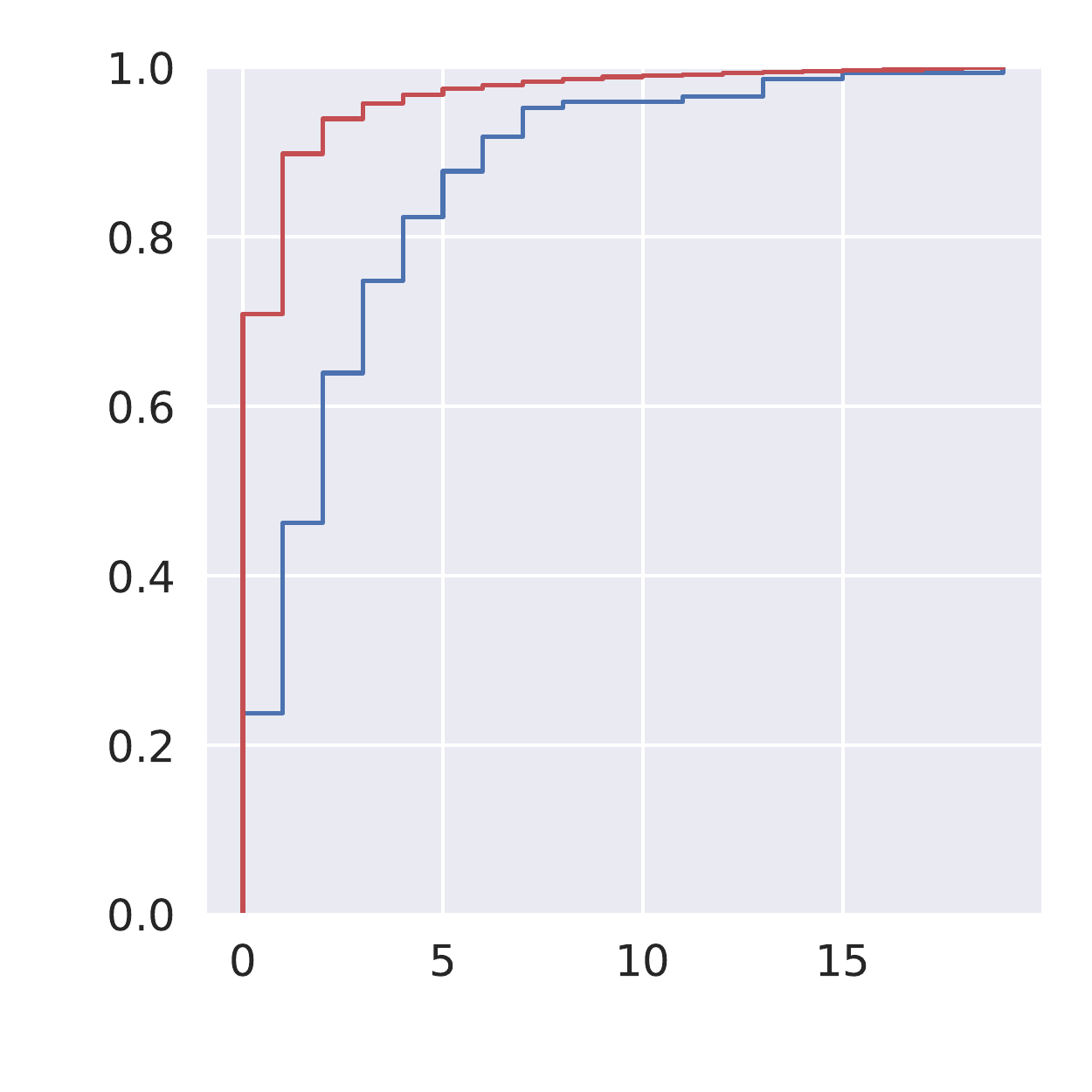}
    }
    \subfigure[\textbf{hi}]{
    \includegraphics[width=0.145\linewidth]{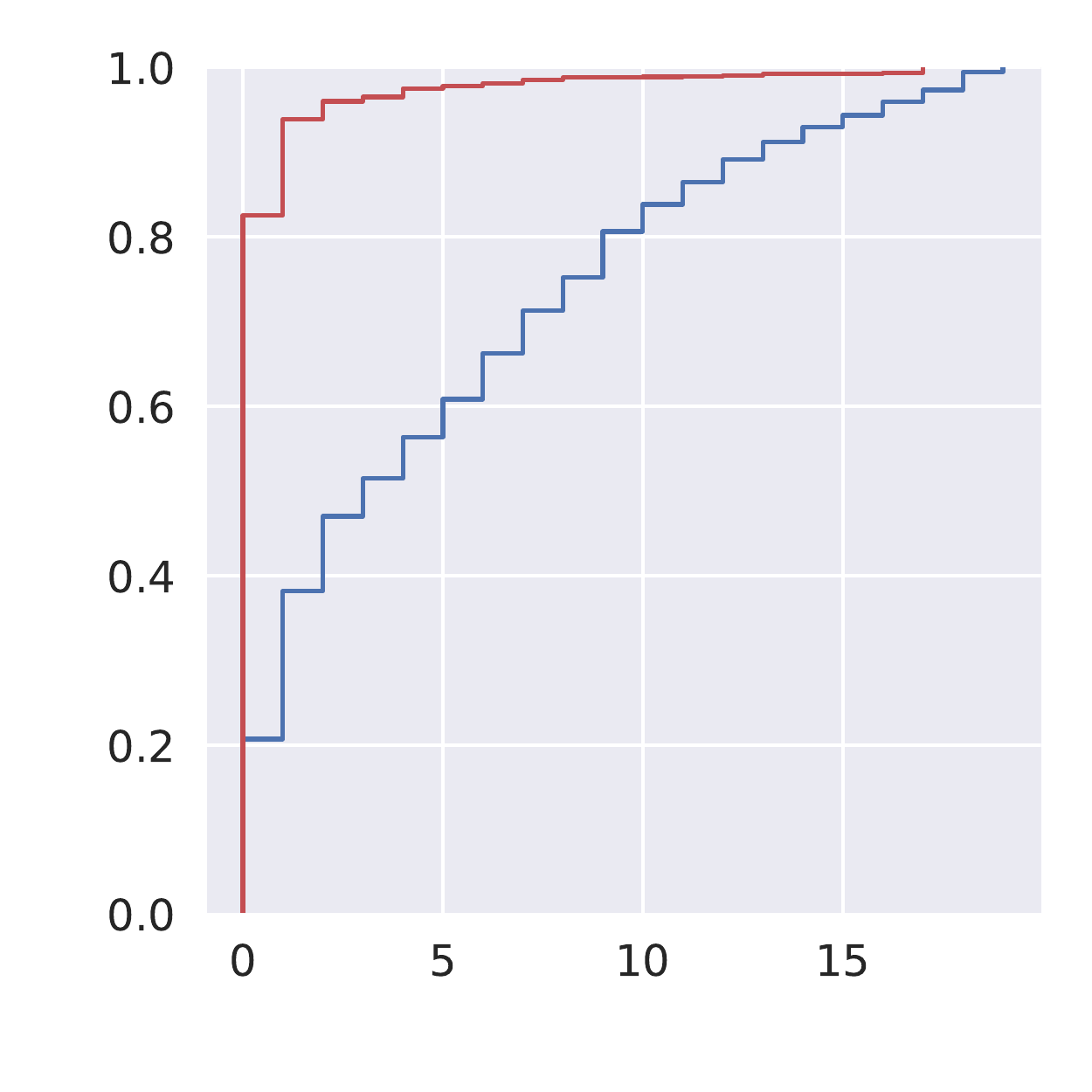}
    }
    \subfigure[\textbf{fr}]{
    \includegraphics[width=0.145\linewidth]{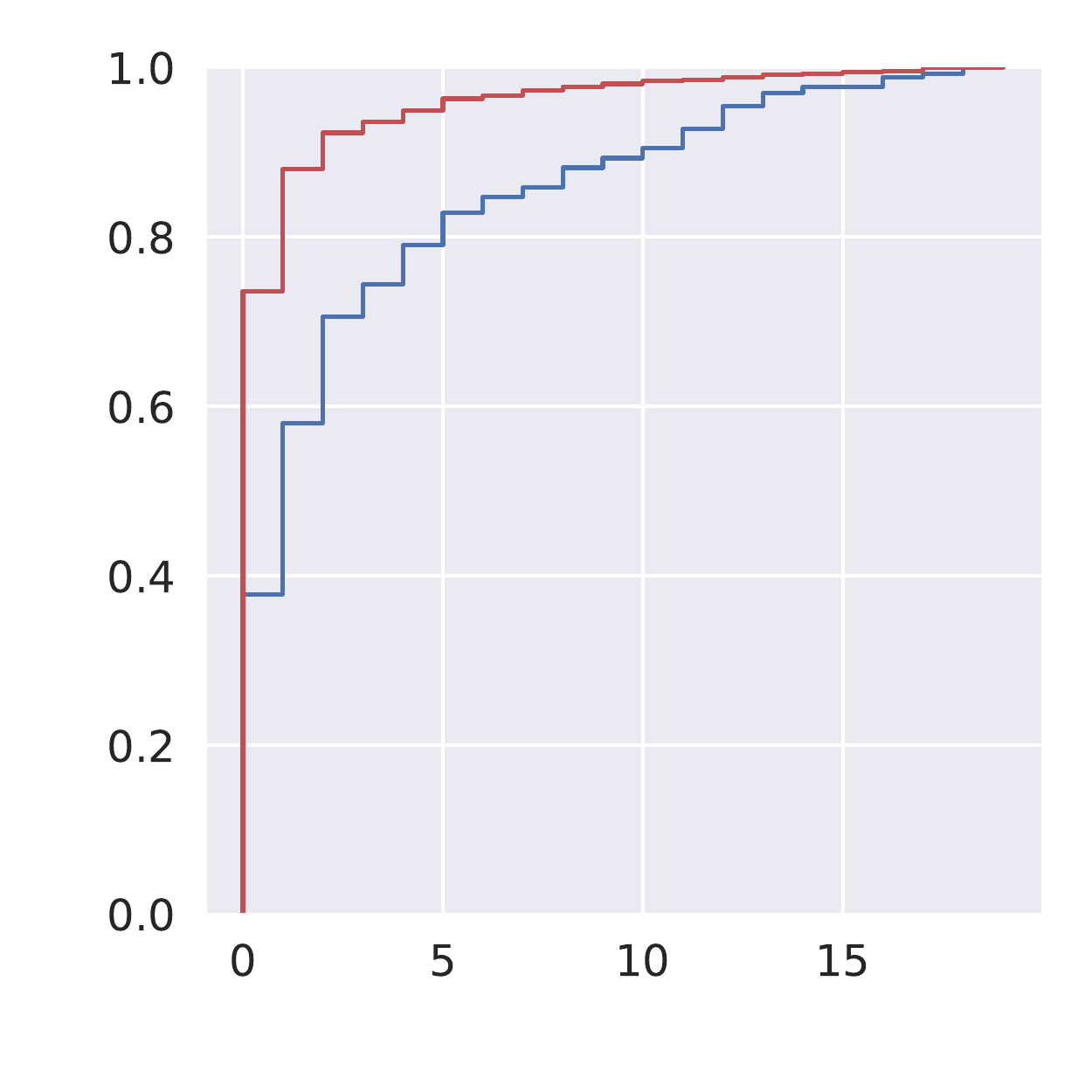}
    }
    \subfigure[\textbf{it}]{
    \includegraphics[width=0.145\linewidth]{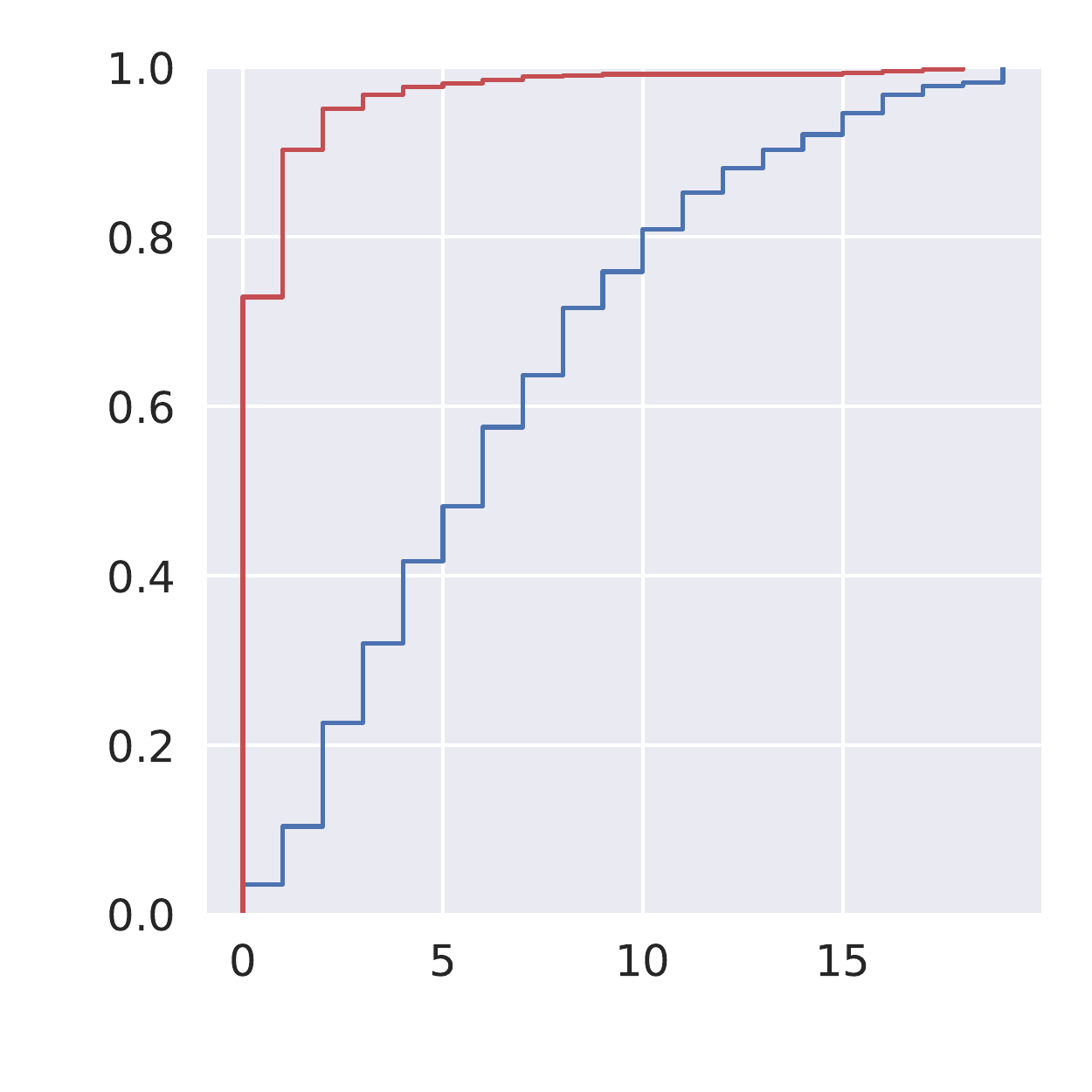}
    }
    
    \caption{The accumulated distribution of number of replies for each language. \textcolor{blue}{Blue} stands for real news related tweets and \textcolor{red}{Red} is for fake news related tweets.}
    \label{fig:replies_distri}
\end{figure*}

\begin{figure*}[!tbh]
    \centering
    % \subfigure[\textbf{All}]{
    % \includegraphics[width=0.13\textwidth]{figure/all_tweets_retweet_count.pdf}
    % }
    \subfigure[\textbf{en}]{
    \includegraphics[width=0.145\textwidth]{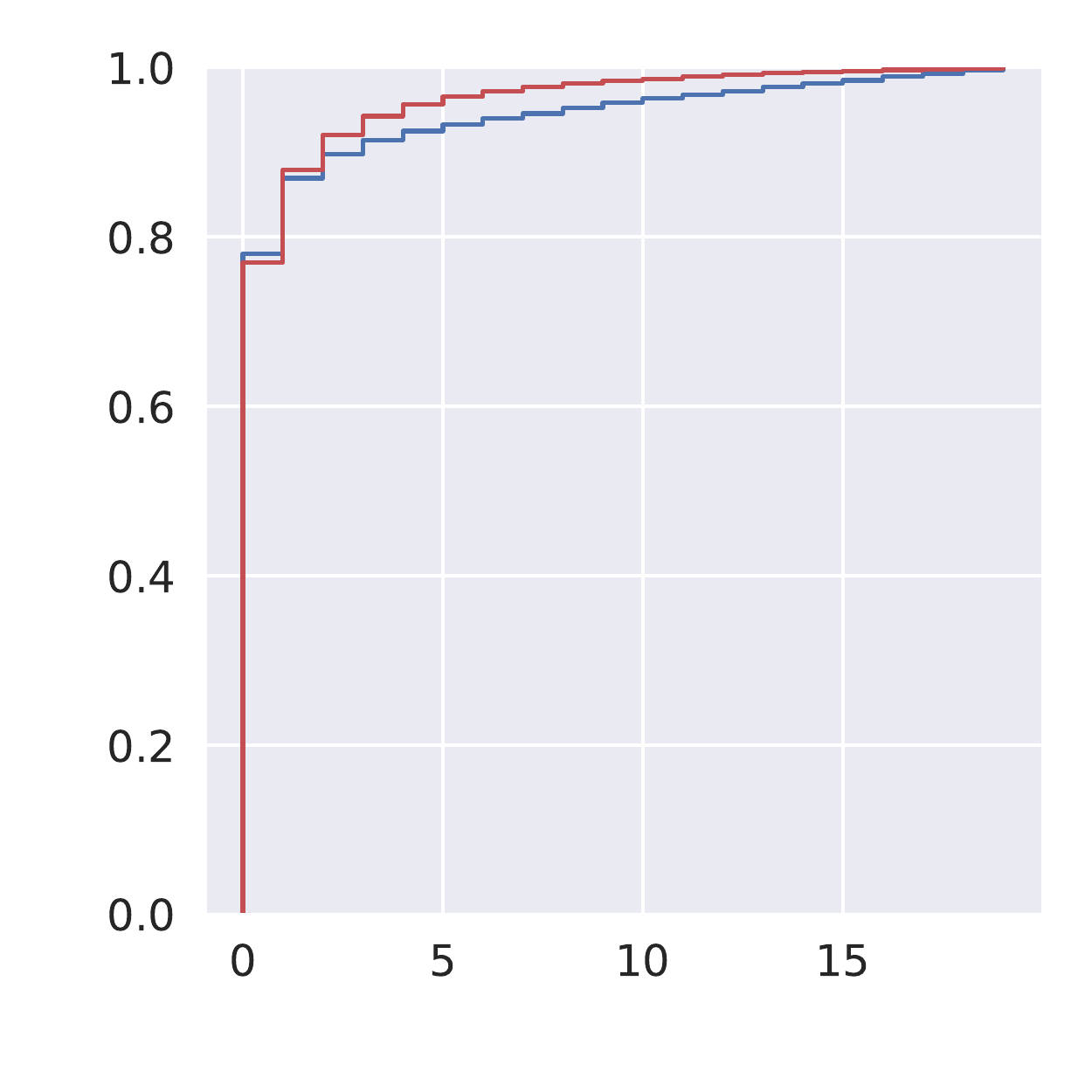}
    }
    \subfigure[\textbf{es}]{
    \includegraphics[width=0.145\textwidth]{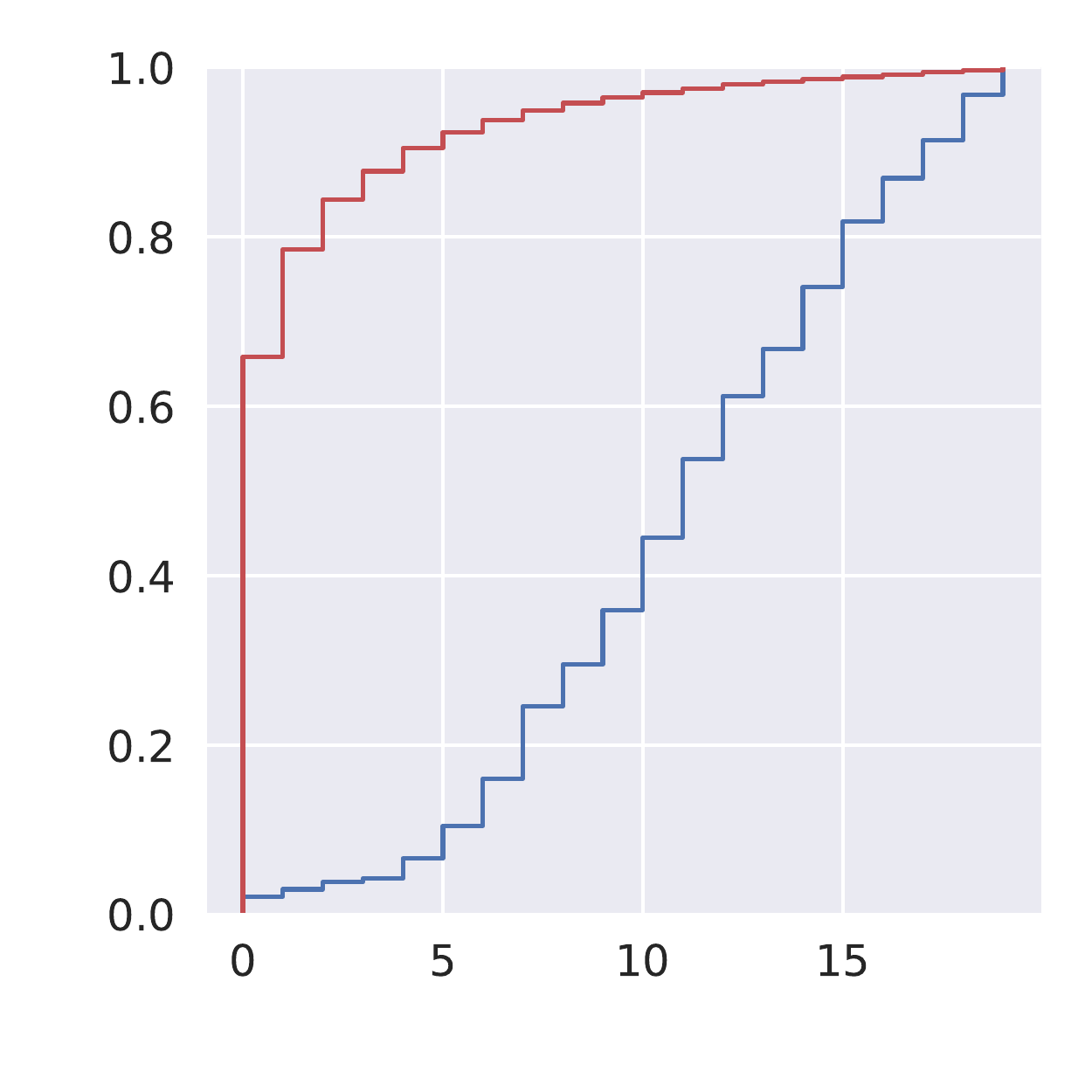}
    }
    \subfigure[\textbf{pt}]{
    \includegraphics[width=0.145\textwidth]{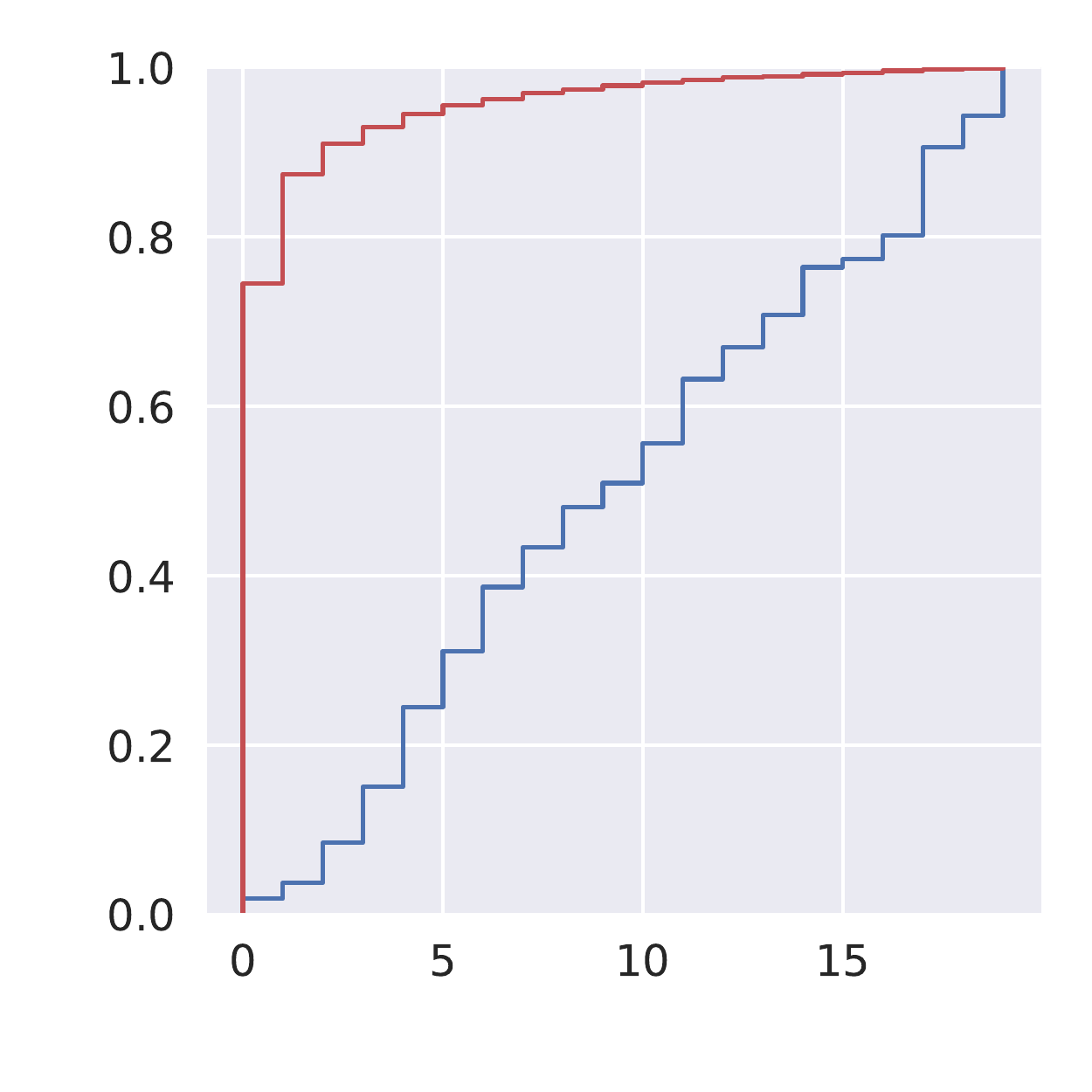}
    }
    \subfigure[\textbf{hi}]{
    \includegraphics[width=0.145\textwidth]{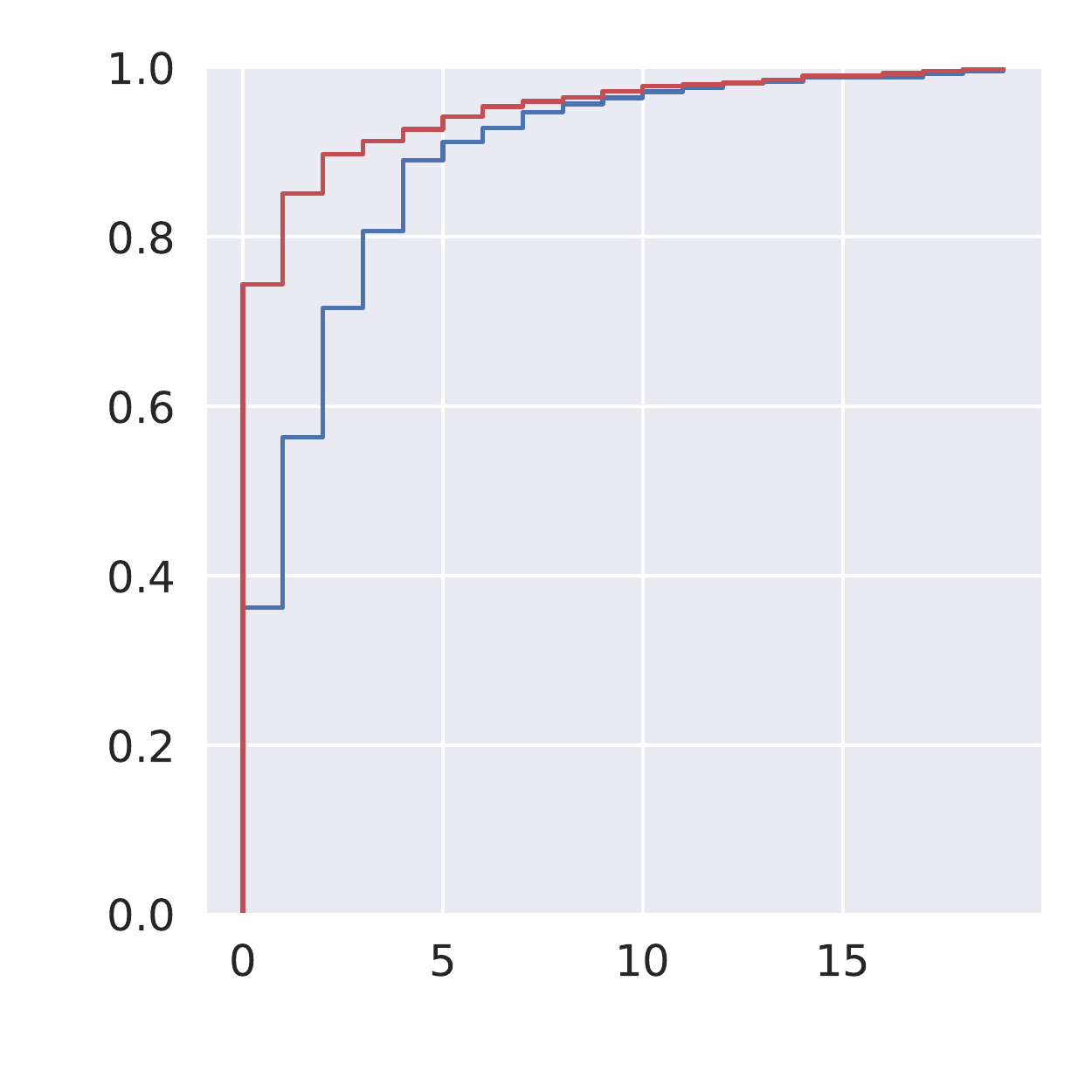}
    }
    \subfigure[\textbf{fr}]{
    \includegraphics[width=0.145\textwidth]{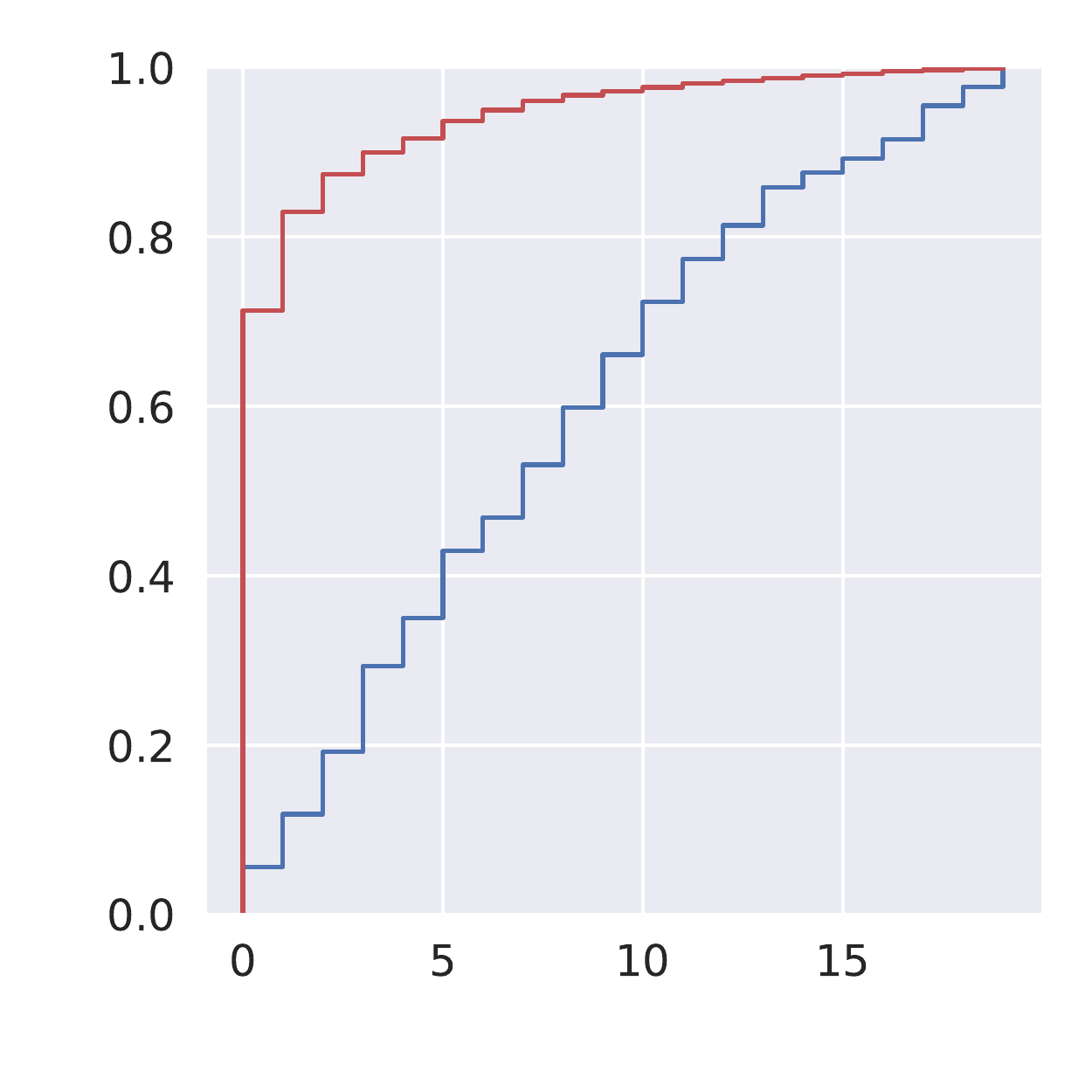}
    }
    \subfigure[\textbf{it}]{
    \includegraphics[width=0.145\textwidth]{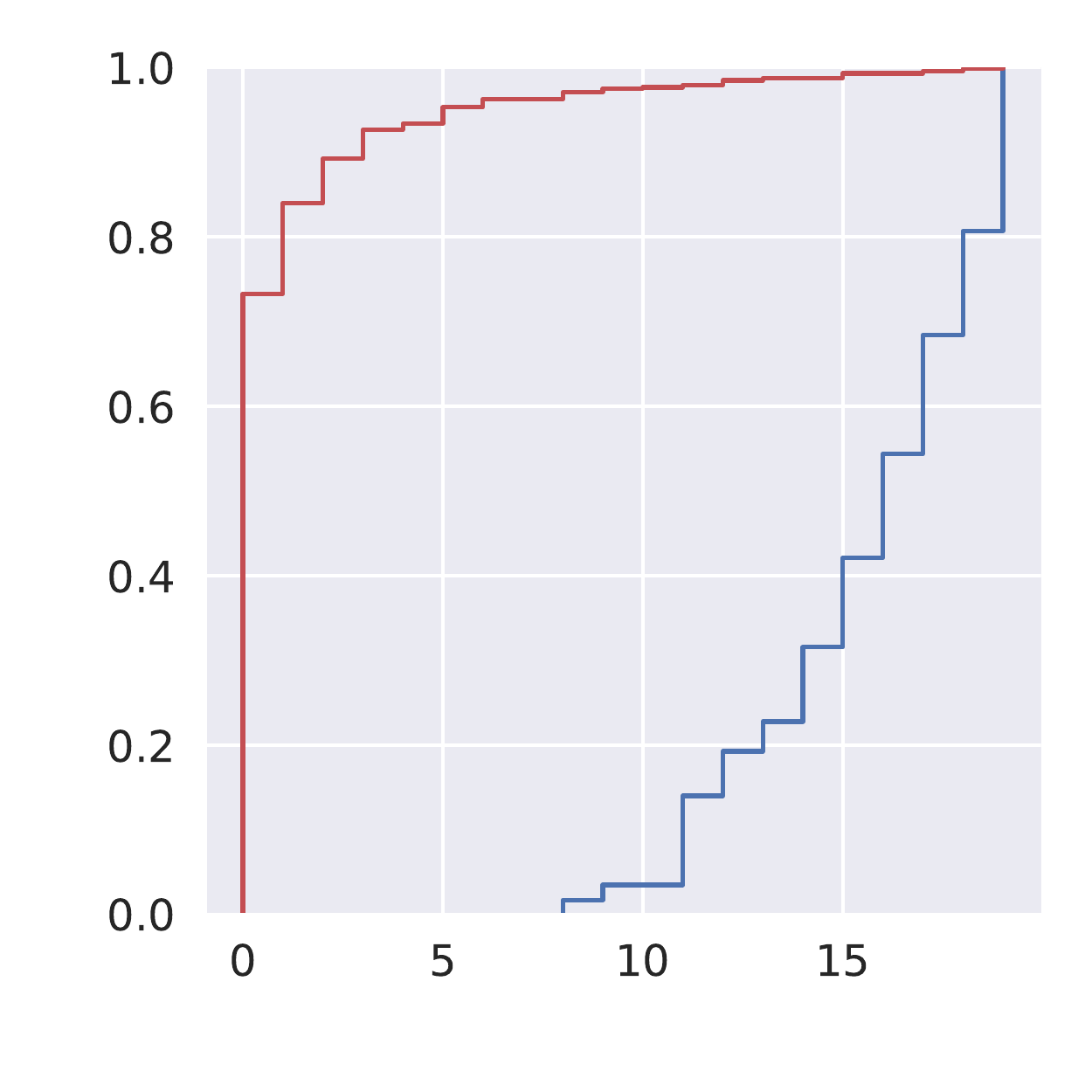}
    }
    \caption{The accumulated distribution of number of retweets for each language. \textcolor{blue}{Blue} stands for real news related tweets and \textcolor{red}{Red} is for fake news related tweets. }
    \label{fig:retweet_distri}
\end{figure*}

To gain insights into user interaction intensify between the fake news and real news, we reveal the distribution of the count of retweets and replies towards them. From Figure~\ref{fig:replies_distri} and Figure~\ref{fig:retweet_distri}, we can find that for languages except \textbf{en} real news get larger number of replies and retweets than the fake news. But in \textbf{en}, there is no significant difference between the real news and the fake news. These observations indicate that language also impacts users' social interactions. 

% One research~\citep{credibility2013news} points out the low credibility users will work coordinated to publicize the content to suppress independent people.

Next, to understand the topic difference between the  tweets of fake and real news, we reveal the most salient hashtags in Table~\ref{tab:twitter_hashtags}. We remove the frequent hashtags like \textit{\#COVID-19, \#Coronavirus, \#sars} to better provide distinct patterns. From Table~\ref{tab:twitter_hashtags}, we observe that there exits consistent difference in several languages. For example, in \textbf{en}, fake news tweets mentioned the key words of common conspiracy theories  like \textit{\#vaccine} and \textit{\#hydroxychloroquine}. This also happens in language like \textbf{fr} and \textbf{it}. fake news of \textbf{fr} mentions \textit{\#5g}, \textit{\#chloroquine} and \textit{\#antimasque}, and \textbf{it} mentions \textbf{\#vaccino}. Besides, fake news tweets in \textbf{en} and \textbf{hi} mention the politic keyword \textit{\#trump} and \textit{\#telanganaliberation} respectively. However, the real news tweets in these languages either focus on official health agency like \textit{\#nih} in \textbf{en} or general exhortation of defending COVID-19 like \textit{\#healthforall} in \textbf{hi}, 
\textit{\#stopthepandemic} and \textit{\#prevention} in \textbf{fr} and  \textit{\#restiamoadistanza} in \textbf{it}. In the meantime, there is no significant topic difference in the \textbf{es} and \textbf{pt}, the fake news and real news both talk about the general exhortation. 
% These observation reveals the challenges of fake news detection cross the languages. 

\begin{table*}[tbh!]
    \centering
    \caption{Salient hashtags in tweets toward fake news and real news.}
    \begin{tabular}{|c|c|J}
         \hline
         Veracity & Language & Hashtags \\ 
         \hline
         \multirow{7}{*}{Fake} 
         
         & en &\#lockdown, \#lka, \#stayhome, \#vaccine, \#hydroxychloroquine, \#trump, \#amitshah \\
         \cline{2-3}
         & es & \#ecuador, \#pandemia, \#salud \#quedateencasa, \#notasdeprensa, \#bolivia \\
         \cline{2-3}
         & pt & \#pandemia, \#auxilioemergencial, \#botucatu, \#brasil, \#agênciasaúde \\
         \cline{2-3}
         & hi & \#telanganaliberation, \#unlocktheopen, \#lockdown, \#socialdistancing, \#healthforall \\
         \cline{2-3}
         & fr & \#5g, \#chloroquine, \#masque, \#antimasque, \#unlocktheopen \\
         \cline{2-3}
         & it & \#notizie, \#vaccino, \#plasma \\
         
         \hline
         \multirow{7}{*}{Real}

         &  en & \#ai, \#artificialintelligence, \#health, \#usa, \#nih, \#technology \\
         \cline{2-3}
         & es & \#quédateencasa, \#estevirusloparamosunidos, \#personaldesalud, \#plasmaconvaleciente  \\
         \cline{2-3}
         & pt & \#estamoson, \#desconfinamento, \#estudoemcasa, \#umconselhodadgs, \#naofacilites \\
         \cline{2-3}
         & hi & \#swasthabharat, \#healthforall, \#lockdown2 \\
         \cline{2-3}
         & fr & \#epidémio, \#osh, \#stopthepandemic, \#prévention, \#surveillance \\
         \cline{2-3}
         & it & \#restiamoadistanza, \#iorestoacasa, \#fase2, \#resistiamoinsieme \\
         \hline
    \end{tabular}
    
    \label{tab:twitter_hashtags}
\end{table*}

% \noindent\textbf{Following Network}

\subsection{Temporal Information}
\begin{figure*}[!tbh]
    \centering
    \includegraphics[width=\linewidth]{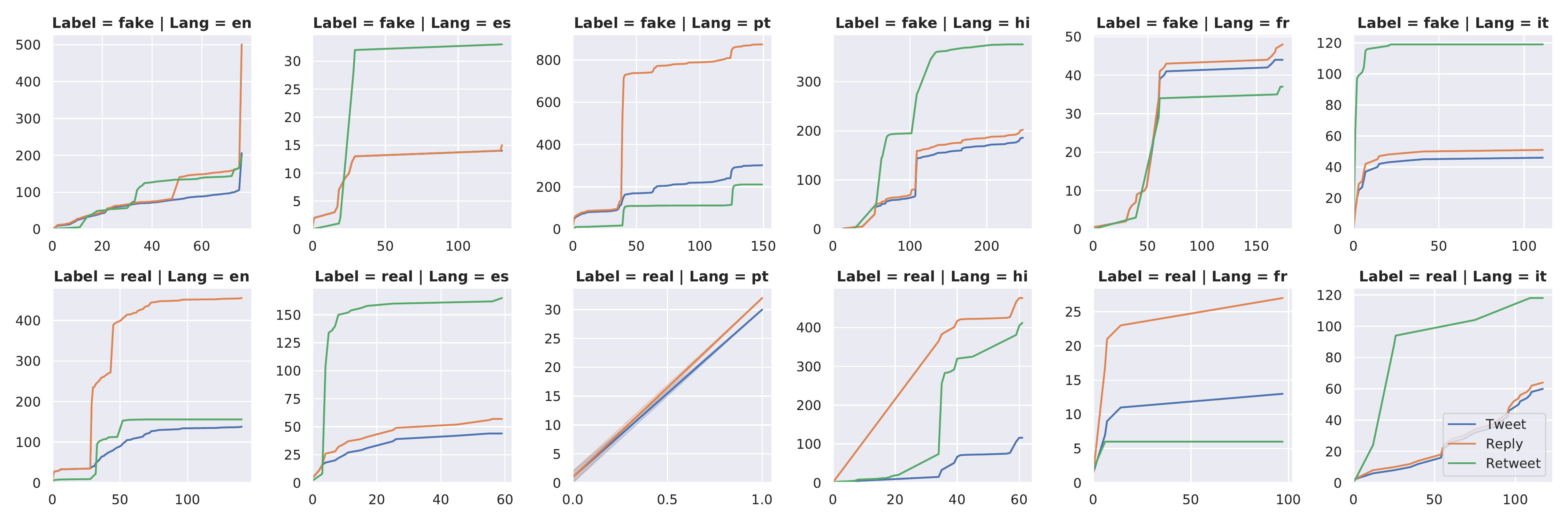}
    \caption{Temporal patterns of social engagement of fake news and real news in different languages.}
    \label{fig:timeline}
\end{figure*}

Recent researches have shown that the temporal information of social engagements can improve fake news detection performance~\cite{temporal, temporal_kai}. To reveal the temporal patterns difference between real news and fake news, we follow the analysis approaches in~\cite{shu2018fakenewsnet, dai2020ginger} that select two news pieces for each language and reveal the count. From Figure~\ref{fig:timeline}, we observe that (i) real news in en, es, pt, hi, and fr have a sudden increase in social engagements. (ii) in the language, on the contrary, there is a steady increase in the real news. These common temporal social engagement patterns allow us to extract the language invariant features for fake news detection.

% Next, it has been suggested that  features of users temporal social engagements can be used to fake news detection\cite{fakenewstracker}\cite{temporal}. To understand the temporal pattern of tweets in {\data}, we select one content for each language and reveal the tweets, replies and retweets time interval in Figure~\ref{fig:reply_time_line}. From Figure~\ref{fig:reply_time_line}, we can find xx. 

\section{fake news Detection}\label{sec:experiment}
In this section, we select several baseline methods to perform fake news detection on {\data}. Since the COVID-19 is the global pandemic, the COVID-19 fake news has been spread all over the world. There are three different stages of fake news spreading in one language: at the beginning, there is no fake news resource(labeled fake news content), in the middle, there is a limited resource and in the end, there is enough resource. We aim to answer three research questions under different resource settings: 
\begin{itemize}
 
    \item \textbf{RQ1 Enough Resource:} what is the fake news classification performance on each language when there is \textbf{enough resource}? 
    \item \textbf{RQ2 Low Resource:} what is the fake news classification performance for each language when there is \textbf{low resource} at that language? 
      \item \textbf{RQ3 No Resource}: what is the fake news classification performance for each language when there is \textbf{no resource} at that language? 
\end{itemize}

\subsection{Baseline Methods}
We deploy several fake news detection methods as following:
\begin{itemize}
    \item \textbf{Text Content:} Models in this group only utilize the fake news text content to do the detection. We apply several classification methods like Support Vector Machine~(SVM), XGBOOST~(XGB), and the variant of dEFEND~\cite{defend}-dEFEND\textbackslash\textbf{C} which utilize sentence attention LSTM model to learn the representation of the news content. 
    
    \item \textbf{Social Context:} The social context-based models utilize the social engagements to do the fake news classification. We utilize the variant of dEFEND~\cite{fakenewstracker}-dEFEND\textbackslash\textbf{N}, which utilizes the user's reply sequences for fake news detection.    
    \item \textbf{Text Content and Social Context:}  dEFEND~\cite{defend}  utilize the fake news reply from the user social engagements and fake news content to do the fake news detection.
    \end{itemize}
% We use 80\% of data for training and 20\% for test for each language. 
\subsection{Implementation Detail}

The overall dataset is randomly divided into training and testing datasets while the proportion is based on the different resource settings. To control counterfactual features of the dataset(the length of fake news, the existence of social engagements), remove the fake news samples whose length is shorter than 10-word tokens, and whose count of replies and tweets is zero. In addition, we balance the fake news and real news. This result in 1,006, 174, 300, 142, 90, and 70 samples in en, es, pt, hi, fr, it respectively. For each method, we repeat the experiment 5 times and report the average accuracy and Macro-F1 score. For traditional machine learning methods~(SVM and XGBoost), we utilize bag-of-words to represent the text. For neural network-based methods(dEFEND and its variants), we utilize XLM-RoBERTa~\cite{conneau2020unsupervised} to get the representation of the text without fine-tuning. 
\subsection{Experimental Results}
To answer these three research questions, we set up three different experiment settings: 
\newline\noindent\textbf{Enough Resource:}\label{sec:general}
We train the fake news classification model on 80\% data and test on the left 20\% data for each language. The experiment result is provided in Table~\ref{tab:ex_result}. We observe that \textit{(i)} for content-based approaches, dEFEND\textbackslash \textbf{C} achieves the best performance and all content baseline methods achieve reasonable performance in all languages; \textit{(ii)} the social-context and content-based method dEFEND achieves the best performance compared with model only utilize the content and social context. These experimental observations indicate the importance of social engagements in fake news detection and the quality of {\data} in each language.  

\begin{table*}[]
    \centering
        \caption{Experiment result of existing fake news detection methods on {\data}}
    \begin{tabular}{|l|cc|cc|cc|cc|cc|cc|}
         \hline
         Language & \multicolumn{2}{c}{en} & \multicolumn{2}{c}{es} & \multicolumn{2}{c}{pt} & \multicolumn{2}{c}{hi} & \multicolumn{2}{c}{fr} & \multicolumn{2}{c}{it} \\
         \hline
         Metric & Acc & F1 & Acc & F1 & Acc & F1   & Acc & F1 & Acc & F1 & Acc & F1 \\
         \hline
         SVM &  0.74 & 0.73	& 0.74 &0.74 &	0.87 & 0.87 & 0.72 & 0.72 &  0.72 & 0.72 & 0.79	&  0.78  \\ 
         XGB & 0.75 & 0.74 & 0.74 &	0.74 &	0.89 & 0.90   & 0.73 & 0.72 & 0.67 & 0.67  & 0.81&	0.8  \\
        %  CNN & 0.88 & 0.88 & 0.79	& 0.78 &  \textbf{0.80} & \textbf{0.80} & 0.75 & 0.73 & 0.82	& 0.81 & \textbf{0.80} & \textbf{0.80}\\
         dEFEND\textbackslash\textbf{C} & 0.77 & 0.77 & 0.76 & 0.75 & {0.91} & {0.91} & {0.86} & {0.86} & \textbf{0.95} & \textbf{0.95} & 0.82 & 0.83 \\
         dEFEND\textbackslash\textbf{N} & 0.82 & 0.82 & {0.89} & {0.86} & 0.90 & 0.89 & 0.93 & 0.93 & 0.84 & 0.84 & \textbf{0.91} & 0.91  \\
         dEFEND & \textbf{0.91} & \textbf{0.90} & \textbf{0.91} & \textbf{0.92} & \textbf{0.95} & \textbf{0.96} & 0.96 & 0.96 & 0.91 & 0.91 & \textbf{0.91} & \textbf{0.92} \\
        %  \midrule 
        %  XGB\_{Social} & & & \\
        %  \midrule
                %  \midrule
        %  XGB\_{Content+Social} & & & \\
        %  TCNN\_URG & 0.86 & 0.81 & 0.91 \\
        %  dEFEND & 0.93 & 0.91 & 0.53 \\
         \hline
    \end{tabular}

    \label{tab:ex_result}
\end{table*}

\noindent\textbf{Low Resource:} there is a limited number of target language resources and enough other language resources. We jointly train the model on multiple source languages and limited target language samples then apply the model to the target language. For each source language, there are 80\% of data for training, and for the target language, there are only 20\% data for training and also 20\% of data for the test.  From the experiment result shown in  Table~\ref{tab:ex_adapt}, we find that \textit{(i)} without any source language, dEFEND achieves the best performance across all languages and dEFEND\textbackslash\textbf{N} achieve better performance than dEFEND\textbackslash\textbf{C} in most languages(en, hi and fr). This indicates that the social context provides the auxiliary information when there is a limited resource; \textit{(ii)} in language es, the additional languages improve the performance in dEFEND\textbackslash\textbf{N} and dEFEND models and in language fr, the additional languages improve the dEFEND\textbackslash\textbf{C} and dEFEND\textbackslash\textbf{N}. However, in other cases, simply combine different languages brings much noise in classification causing worse performance.

\begin{table*}[]
    \centering
     \caption{Experiment result of \textbf{low-resource} on {\data}.  The source domains are the languages except the target domain. }
    \begin{tabular}{|l|cc|cc|cc|cc|cc|cc|}
         \hline
         Language & \multicolumn{2}{c|}{en} & \multicolumn{2}{c|}{es} & \multicolumn{2}{c|}{pt} & \multicolumn{2}{c|}{hi} & \multicolumn{2}{c|}{fr} & \multicolumn{2}{c|}{it} \\
         \hline
         Metric & Acc & F1 & Acc & F1 & Acc & F1   & Acc & F1 & Acc & F1 & Acc & F1 \\
         \hline
        %  dEFEND\textbackslash\textbf{C} Trans &  & 0.88 & 0.79	& 0.78 &  \textbf{0.80} & \textbf{0.80} & 0.75 & 0.73 & 0.82	& 0.81 & \textbf{0.80} & \textbf{0.80}\\
        \multicolumn{13}{l}{\textbf{Without Additional Language Dataset}} \\ 
        \hline
         dEFEND\textbackslash\textbf{C} & 0.78	& 0.78 & 0.64 & 0.61 & 0.79 & 0.79 & 0.84 & 0.83 & 0.76 & 0.75 & 0.85 & 0.85 \\
         dEFEND\textbackslash\textbf{N}  & 0.77 & 0.77 &	0.77 & 0.77  & 0.84 & 0.84 & 0.84 & 0.85 & {0.78} & 0.77  & 0.79 & 0.78 \\
         dEFEND & \textbf{\underline{0.85}} & \textbf{\underline{0.85}} & \underline{0.76} & \underline{0.75}  & \textbf{\underline{0.90}} & \textbf{\underline{0.90}} & \textbf{\underline{0.90}} & \textbf{\underline{0.90}} & \textbf{\underline{0.89}} & \textbf{\underline{0.89}} & \textbf{\underline{0.86}} & \textbf{\underline{0.85}} \\
         \hline
         \multicolumn{13}{l}{\textbf{With Additional Language Dataset}} \\
         \hline
         dEFEND\textbackslash\textbf{C} & 0.67 & 0.64 & 0.63 & 0.61 &  0.75 & 0.75 & 0.79 & {0.78} & 0.84 & 0.84 & \underline{0.85} & \underline{0.85}\\
         
        %  dEFEND\textbackslash\textbf{N} Trans \\
         dEFEND\textbackslash\textbf{N} & 0.71 & 0.71 &  \underline{\textbf{0.87}} & \underline{\textbf{0.87}} & \underline{0.85} & \underline{0.85} & \underline{0.84} & \underline{0.84} & 0.87 & 0.86 & 0.74 & 0.74 \\
         
        %  dEFEND Trans & 0.64 & 0.62 & \textbf{0.89} & \textbf{0.89}  & \textbf{0.80} & 0.78 & \textbf{0.83} & \textbf{0.82} & \textbf{0.83} & \textbf{0.83} & \textbf{0.80} & 0.78\\
         dEFEND  & \underline{0.82} & \underline{0.81}  & 0.77	& 0.77 & 0.83 & 0.83 & 0.78 & 0.77 & \underline{{0.88}} & \underline{{0.88}} & 0.81 & 0.79 \\
        %  \midrule 
        %  XGB\_{Social} & & & \\
        %  \midrule
                %  \midrule
        %  XGB\_{Content+Social} & & & \\
        %  TCNN\_URG & 0.86 & 0.81 & 0.91 \\
        %  dEFEND & 0.93 & 0.91 & 0.53 \\
         \hline
    \end{tabular}
   
    \label{tab:ex_adapt}
\end{table*}

\noindent\textbf{No Resource:} there are no resources for the target language. This situation is that fake news spread in a new language, there is no labeled fake news content in this specific language to train the language-dependent fake news detection model. For each language, we split the dataset into 80\% and 20\% for training and test respectively. For simplicity, we only train the detection model in one source language then apply it to the target language. From the experiment result shown in Table~\ref{tab:cross_language}, we can observe that social information plays an important roles in most languages(en, es, pt, hi, and fr; dEFEND\textbackslash\textbf{N}, dEFEND $>$ dEFEND\textbackslash\textbf{C}), this experiment result indicates that social context can provide the language invariant features for the cross-lingual fake news detection. 
\begin{table}[h!]
    \centering
        \caption{Baseline methods classification accuracy for \textbf{no-resource} setting on the test of {\data}. Underlined scores indicate the best result on each transfer language for each group and bold scores the overall best accuracy. }
    \begin{tabular}{|c|cccccc|}
    \hline
    Training & \multicolumn{6}{c|}{Test Language} \\
    Language & en &  es & pt & hi & fr & it \\
    \hline
    \multicolumn{2}{l}{\textbf{dEFEND\textbackslash\textbf{C}}}  \\
    \hline
    en &  (0.77) & \underline{0.61} & 0.50 &  \underline{0.57} & {0.68} & \textbf{\underline{0.73}}\\ 
    es & 0.50 & (0.76) & 0.47 & 0.56 & 0.54 & 0.54 \\
    pt & 0.50 & \underline{0.61} & (0.91) & 0.48 & 0.50 & 0.50 \\
    hi & \underline{0.60} & 0.57 & \underline{0.52} & (0.86) & 0.80 & {0.51}  \\
    fr & 0.53 & 0.54 & 0.46 & 0.51  & (0.95) & 0.54 \\
    it & {0.59} & 0.56 & {0.47} & 0.44 & \underline{0.82} & (0.83) \\
    \hline
    
    \multicolumn{2}{l}{\textbf{dEFEND\textbackslash\textbf{N}}} \\ 
    \hline
    en & (0.83) &  \underline{0.60} & 0.63 & \textbf{\underline{0.76}} &  {0.60} & {0.41} \\
    es & 0.50 & (0.89) & \textbf{\underline{0.67}} & 0.53 & 0.58 & 0.53\\
    pt & 0.57 & {0.57} & (0.89) & 0.48 & 0.63 & 0.49 \\
    hi & \textbf{\underline{0.80}} & {0.56} & {0.50} & (0.93) & {\underline{0.67}} & 0.36 \\
    fr & {{0.70}} & 0.57 & {0.58} & 0.71 & (0.84) & \underline{{0.57}} \\
    it & 0.50 & 0.49 & 0.50 & 0.50 & 0.50 & (0.91)\\
    \hline
    \multicolumn{2}{l}{\textbf{dEFEND}} \\
    \hline
    en & (0.91) & 0.52 & 0.50 & {0.64} & 0.68 & 0.50\\
    es & {0.52} & (0.91) & \underline{0.57} & 0.65 &  {0.68} & {0.50}\\
    pt & 0.50 & 0.49 & (0.96) & 0.50 &  0.51  & 0.47 \\
    hi & 0.62 & 0.55 & {{0.51}} & (0.96) &  \underline{\textbf{0.83}}  &  0.50\\
    fr & \underline{0.65} & \underline{\textbf{0.63}} &  0.48 & {\underline{0.68}} & (0.91) & {\underline{0.60}}\\
    it & 0.51 & {0.46} &  0.50 &  0.48 &  0.57 & (0.91)\\
    \hline
    \end{tabular}

    \label{tab:cross_language}
\end{table}

% \begin{table*}[]
%     \centering
%     \begin{tabular}{l|cc|cc|cc|cc|cc}
%          \toprule
%          Language &  \multicolumn{2}{c}{es} & \multicolumn{2}{c}{pt} & \multicolumn{2}{c}{hi} & \multicolumn{2}{c}{fr} & \multicolumn{2}{c}{it} \\
%          \midrule
%          Metric & Acc & F1 & Acc & F1   & Acc & F1 & Acc & F1 & Acc & F1 \\
%          \midrule
         
%          dEFEND\textbackslash\textbf{C} & 0.76 & 0.76  &  0.47 & 0.42 & 0.75 & 0.75 & x& x & 0.60 & 0.58\\
%         dEFEND\textbackslash\textbf{N} & 0.71 & 0.71 & 0.57 & 0.51 & 0.58 & 0.56 & x& x & 0.60 & 0.58\\
%          Defend & 0.76 & 0.76 & 0.43 & 0.35 & 0.83 & 0.83&x & x & 0.60	& 0.58\\
%         %  \midrule 
%         %  XGB\_{Social} & & & \\
%         %  \midrule
%                 %  \midrule
%         %  XGB\_{Content+Social} & & & \\
%         %  TCNN\_URG & 0.86 & 0.81 & 0.91 \\
%         %  dEFEND & 0.93 & 0.91 & 0.53 \\
%          \bottomrule
%     \end{tabular}
%     \caption{Experiment result of language adaptation methods on {\data}.  The source domain are the English and the target domain are the rest language. }
%     \label{tab:ex_single}
% \end{table*}

\section{Potential Application}\label{sec:insight}
Our goal is to provide a comprehensive COVID-19 fake news dataset to help research around COVID-19 infodemic. This dataset provide multilingual and multi-modal information which could benefit in various topics like cross-lingual and early fake news detection; fake news propagation and fake news mitigation. 

\subsection{Fake News Detection}
Our goal is to provide a comprehensive COVID-19 fake news dataset to help researchers around the COVID-19 infodemic. This dataset provides multilingual and multi-modal information that could benefit from various topics like cross-lingual and early fake news detection; fake news propagation and fake news mitigation. 

\noindent{\textbf{Cross-Lingual Fake News Detection:}} The multilingual characteristics bring two new applications from a language perspective. On the one hand, with the daily emerging COVID-19 fake news, we can correlate the knowledge we learn from different languages to improve the overall fake news detection performance for the future; and on the other hand, for languages that are poor with annotated fact-checking labels, we can transfer the knowledge in rich source languages such as English towards these low resource languages. The past cross-lingual research like abusive language detection~\cite{pamungkas-patti-2019-cross}, cross-lingual rumor verification~\cite{wen-etal-2018-cross} and cross-lingual hate speech detection~\cite{stappen2020crosslingual} have shown proven performance in either languages cooperation or low resource language. These approaches only utilize the text information through extracting the language invariant features and encoding the text content into a shared embedding space to achieve knowledge transferring among different languages. Since fake news is intentionally written to misled audiences, the approaches of only utilizing the content in a monolingual setting are hard~\cite{shu2018fakenewsnet}, let alone cross-lingual. Our dataset provides auxiliary information like social engagements. dEFEND~\cite{defend} integrate the users' replies into fake news representation learning and Shao~\cite{social_bots} propose a method utilize the user profile into fake news detection. Thus, {\data} provides a comprehensive dataset to study the cross-lingual fake news detection by expanding the feature space including the fake news content and social engagements.

\noindent{\textbf{Early Fake News Detection:}} The COVID-19 fake news has already brought uncertainty, fear, and racism globally. To defend future epidemic fake news and resolve the impacts of the fake news, it is urgent to identify the fake news at the early stage before it was widely spread~\cite{early_detection}. This indicates that there is limited social engagements can be used for detection. Our dataset contains the timestamp for the engaged tweets, retweets, and replies which allow researchers to set specific early time windows to understanding the pattern difference between the fake news and real news. Besides, user characteristic plays a very important role in early fake news detection~\cite{shu2019role}. We include user profiles, recent timelines and follower-friend networks in {\data} where we can extract useful features and develop early detection models. Overall, this dataset not just provides all the required features but also the flexibility for researchers to do the early fake news detection analysis to defend the next new epidemic. 

\noindent\textbf{Multi-Modal Fake News Detection:} Some of the COVID-19 fake news contents contain figure or video and text in the same time\footnote{https://archive.is/IDzMF\#selection-732.0-732.2}. The existing researches also have suggested that combining the textual and visual features can improve the performance of fake news detection~\cite{Multimodal, shu2017fake, eann}. {\data} contains multi-modal information by keeping the referenced URLs of the pictures and videos embedded in the fake news content. In this way, researchers can develop new models to extract textual and visual features for the COVID-19 fake news detection. 

\noindent\textbf{Cross-Domain Fake News Detection:} {\data} is the mixture of different fake news domains, like political, entertaining, and healthy.
It can help researchers to learn the domain invariant and domain-dependent features for cross-domain fake news detection.

% This domain variety allow researchers to analysis the similarity and difference cross domains and develop domain specific or domain consistent fake news detection methods.
\subsection{Fake News Mitigation}
To overcome the negative impacts of fake news after it was posted, it is urgent to reduce the spread of fake news. The fake news on social media is widely distributed by users' social networks and personalized recommendation algorithm~\cite{bhattacharjee2020disinformation}.  

% It is to overcome the negative impacts of fake news through  the fake news propagation network intervention, content intervention, susceptible users testing~\cite{bhattacharjee2020fake news}. Since {\data} can provide the propagation network in social media, the corrected information from fact checking websites and detailed user profile, it can help researcher to build models to minimize the influence of the fake news. 

\noindent\textbf{Propagation Network Intervention:} 
The aim of propagation network intervention is to prevent the spread of fake news. There are two main approaches~\cite{shu2018fakenewsnet}: (i) \textit{Influence Minimization}: slowing down the spread of fake news during the dissemination process. Past researches~\cite{Limiting, Spread} proposes methods to delete a small set of users in the propagation network to reduce the spread of fake news. (ii) \textit{Mitigation Campaign}: maximizing the spread of true news to combat the dissemination of fake news. Researches in \cite{he2011influence, Limiting, sharma2019combating} select $k$ seed users for true news cascade in the presence of fake news to minimize the users who will be influenced by fake news. 
{\data} can provide rich propagation network information like multiple dissemination paths(tweet, reply, and retweet), and detailed meta information of the interacted users and transmit information which can help researchers to build up heterogeneous diffusion network to assist the understanding of fake news influence minimization and real news influence maximization. 

\noindent\textbf{Personalized Recommendation Algorithm Intervention:}
Since people react more extremely and engage more towards the fake news content, the recommendation algorithm in social media platform will propagate the fake news to attract more users~\cite{bhattacharjee2020disinformation}. The {\data} contains the fake news page and its relevant authorized evidence pages from fact-checking websites. These web pages can help the researchers to develop fake news aware recommendation algorithms to drop the fake news pages. In addition, {\data} provide the users profile metadata and historical tweets which can facilitate the study of personalized fake news aware recommendation algorithm.

\subsection{Fact-checking Accessory:}
Fact-checking accessory aims to improve the efficiency of the debunking process for fact-checking agencies like Snopes and PolitiFact. The manually fact-checking process requires the fact-checkers to not only provide the veracity of the content but also provide additional evidence and context from authorized sources to support their decisions. To fully utilize fact-checkers' professionalism and help them engage with their familiar domains, researchers can build a model to recommend interested suspicious claims to the professional fact-checkers. In addition, it is possible to automatically retrieve evidence content during the fact-checking process. {\data} can provide the metadata of fact-checking reviews with the suspicious claim and the name of fact-checker and the detailed content of the fact-checking reviews. 
This rich information can help the researchers to develop semi-automatic or automatic fact-checking accessories to help the fact-checkers report the fake news.

\section{Conclusion and Future Work}\label{sec:conclude}
To combat the global infodemic, we release a multilingual fake news dataset~{\data}, which contains the news content, social context, and spatiotemporal information in English, Spanish, Portuguese, Hindi, French, and Italian six different languages. Through our exploratory analysis, we identify several languages invariant and language variant features for fake news detection. The experiment result of several fake news detection methods under three different experiment settings (enough, low, and no resource) demonstrate the utility of {\data}. This dataset can facilitate further research in fake news detection, fake news mitigation, and fact-checking efficiency improvement.

There are several potential improvements for future work: (1) include more languages in the dataset, such as Chinese, Russian, Germany, and Japanese. (2) collect social context from different social platforms like Reddit, Facebook, YouTube, and Instagram, and so on. 

\bibliography{aaai}{}
\bibliographystyle{IEEEtran}

\newpage
\section{Appendix}
\subsection{Reliable Information Source:}

The sources of real news are listed in Table~\ref{tab:detail_news}.
\begin{table}[tbh]
    \centering
    \caption{The sources of the real news. The name starting with "@" is the screen name of a Twitter Account. }
    \begin{tabular}{|c|L|}
        \hline
        en & @WHO, @trvrb, @MayoClinic, @NIH, https://www.cdc.gov, https://newsroom.clevelandclinic.org, https://www.ecdc.europa.eu, https://www.healthline.com, https://www.medicalnewstoday.com, https://www.webmd.com, https://www.who.int\\
        \hline
        es & @spanish, @SSalud\_mx, @sanidadgob, https://coronavirus.gob.mx, https://www.mscbs.gob.es, https://ec.europa.eu/info/index\_es\\
        \hline
        pt & @portuguese, @govpt, https://coronavirus.saude.gov.br, https://www.portugal.gov.pt\\
        \hline
        hi & @MoHFW\_INDIA, @CovidIndiaSeva \\
        \hline
        fr & @santeprevention, @eu\_osha,  @EU\_commission, http://www.santepubliquefrance.fr, https://osha.europa.eu/fr, https://ec.europa.eu/info/index\_fr\\
        \hline
        it & @MinisteroSalute, http://www.salute.gov.it/nuovocoronavirus, https://ec.europa.eu/info/index\_it \\
        \hline
    \end{tabular}
    
    \label{tab:detail_news}
\end{table}

\end{document}